\documentclass[fleqn,useAMS,usenatbib]{mnras}

\usepackage{psfrag,color,epsfig}

\usepackage{graphicx}	% Including figure files
\usepackage{amsmath}	% Advanced maths commands
\usepackage{amssymb}	% Extra maths symbols
\usepackage{multicol}	% Multi-column entries in tables
\usepackage{bm}		% Bold maths symbols, including upright Greek
\usepackage{pdflscape}	% Landscape pages
\usepackage{multirow}
\usepackage{tabularx}  
\def\HI{\ion{H}{I}~}

\def\xb{\bar{x}_{\rm \ion{H}{I}}}
\def\Tb{{T_{\rm b}}}

\def\TTb{\tilde{T}_{\rm b2}}
\def\kk{{\bm{k}}}

\def\U{{\bm{U}}}
\def\dd{{\bm{d}}}
\def\thetavec{{\bm{\theta}}}
\def\cl{{\mathcal C}_{\ell}}
\def\cc{\mathcal C}
\def\n{\hat{\bm{n}}}

\def\cov{\bm{X}^{\ell_{\rm i}}}

\newcommand{\be}{\begin{equation}}
\newcommand{\ba}{\begin{eqnarray}}
\newcommand{\ee}{\end{equation}}
\newcommand{\ea}{\end{eqnarray}}

\def\gtsima{$\; \buildrel > \over \sim \;$}
\def\ltsima{$\; \buildrel < \over \sim \;$}
\def\simgt{\lower.5ex\hbox{\gtsima}}
\def\simlt{\lower.5ex\hbox{\ltsima}}
\def\simpr{\lower.5ex\hbox{\prosima}}

\def\msun{{M_\odot}}

\newcommand\tab[1][1.5cm]{\hspace*{#1}}
\usepackage[T1]{fontenc}
\usepackage{ae,aecompl}

\usepackage{newtxtext,newtxmath}
\title[Predictions for 21-cm MAPS]{Predictions for measuring the 21-cm multi-frequency angular power spectrum using SKA-Low}

\author[R. Mondal et al.]{Rajesh Mondal,$^1$\thanks{E-mail: 
\href{mailto:Rajesh.Mondal@sussex.ac.uk}{Rajesh.Mondal@sussex.ac.uk}}  
Abinash Kumar Shaw,$^{2}$ Ilian T. Iliev,$^1$ Somnath Bharadwaj,$^{2}$
\newauthor Kanan K. Datta,$^3$ Suman Majumdar,$^{4,5}$ Anjan K. Sarkar$^6$ and Keri L. Dixon$^7$\\
$^1$ Astronomy Centre, Department of Physics and Astronomy, 
University of Sussex, Brighton BN19QH, UK\\
$^2$ Department of Physics \& Centre for Theoretical Studies, Indian 
Institute of Technology Kharagpur, Kharagpur 721302, India\\
$^3$ Department of Physics, Presidency University, 86/1 College
Street, Kolkata 700073, India\\
$^4$ Discipline of Astronomy, Astrophysics and Space Engineering, Indian Institute of Technology Indore, Simrol, Indore 453552, India\\
$^5$Department of Physics, Blackett Laboratory, Imperial College, London SW7 2AZ, UK\\
$^6$ Astronomy \& Astrophysics Group, Raman Research Institute, Bengaluru 560080, India\\
$^7$ New York University Abu Dhabi, PO Box 129188, Saadiyat Island, Abu Dhabi, United Arab Emirates} 

\date{Accepted 2020 April 09. Received 2020 April 06; in original form 2019 October 11}
\pubyear{2019}

\begin{document}
\label{firstpage}
\pagerange{\pageref{firstpage}--\pageref{lastpage}}
\maketitle

%--------------------------------------------------------------------

\begin{abstract}
The light-cone~(LC) effect causes the mean as well as the statistical 
properties of the redshifted 21-cm signal $\Tb (\n, \nu)$ to change with 
frequency $\nu$ (or cosmic time). Consequently, the statistical 
homogeneity (ergodicity) of the signal along the line of sight~(LoS) 
direction is broken. This is a severe problem particularly during the Epoch 
of Reionization~(EoR) when the mean neutral hydrogen fraction ($\xb$) changes 
rapidly as the Universe evolves. This will also pose complications for large
bandwidth observations. These effects imply that the 3D power spectrum 
$P(k)$ fails to quantify the entire second-order statistics of the 
signal as it assumes the signal to be ergodic and periodic along the 
LoS. As a proper alternative to $P(k)$, we use the multi-frequency 
angular power spectrum~(MAPS) $\cl(\nu_1,\nu_2)$ which does not assume the 
signal to be ergodic and periodic along the LoS. Here, we study the prospects 
for measuring the EoR 21-cm MAPS using future observations with the upcoming 
SKA-Low. Ignoring any contribution from the foregrounds, we find that the EoR 21-cm MAPS can be measured at a confidence level 
$\ge 5\sigma$ at angular scales $\ell \sim 1300$ for total observation time 
$t_{\rm obs} \ge 128$\,hrs across $\sim 44$~MHz observational bandwidth. We also
quantitatively address the effects of foregrounds on MAPS detectability forecast by avoiding signal contained within the foreground wedge in $(\kk_\perp, k_\parallel)$ plane. These results are very relevant for the upcoming large bandwidth EoR
experiments as previous predictions were all restricted to individually analyzing the signal 
over small frequency (or equivalently redshift) intervals.
\end{abstract}

%Incorporating the foreground `wedge', we have also attempted to quantitatively address the effects of foregrounds on MAPS detectability forecast


%--------------------------------------------------------------------

\section{Introduction}
\label{sec:intro}
The Epoch of Reionization (EoR) is one of the important periods in the 
evolutionary history of our Universe. During this epoch, the ionizing 
radiation from the first luminous sources in the Universe gradually 
ionizes the neutral Hydrogen~(\ion{H}{I}) in the intergalactic medium 
(IGM). As more and more of these sources form, the ionized~(\ion{H}{II}) 
regions grow and eventually overlap and fill almost the entire IGM. Our 
present knowledge about this epoch is very limited. The current 
measurements of the Thomson scattering optical depth 
\citep{planck2016,planck_tau16}, a measure of the line of sight (LoS) 
free electron opacity to cosmic microwave background (CMB) radiation in 
the IGM, suggest that the mean neutral fraction $\xb$ falls by 
$\sim 0.1$ at $z \sim 10$ from a completely neutral IGM. The second 
observation is the Gunn-Peterson optical depth of the high redshift 
quasar spectra \citep{becker01,fan06,fan02,becker15}. These measurements 
show an absorption trough at $z \la 6$, which indicates that the IGM 
was neutral at $0.1$~per~cent level by $z \sim 6$. The third and the most 
recent constraint comes from the measurements of the luminosity function 
and clustering properties of high-$z$ Lyman-$\alpha$ emitters 
\citep{konno14,santos16,zheng17,ota17}. These studies indicate a patchy 
distribution of \HI and infer a sharp increase in $\xb$ at redshifts 
larger than $z\sim 7$. The findings of all these indirect observations 
provide an overall indication that the EoR probably extends over a 
redshift range $6 \la z \la 12$ 
\citep{robertson13,robertson15,mitra2015,mitra2017,dai18}. 
However, these indirect observations are not able to shed light on 
various fundamental issues, such as the exact duration and timing of reionization, properties of the ionizing sources, the topology of \HI at 
different cosmic times, etc.

Observations of the redshifted 21-cm signal caused by the hyperfine 
transition of \HI in the IGM is the most promising probe of the EoR 
\citep{scott_1990,Bharadwaj2001}. There has been a considerable 
observational effort devoted to measuring the EoR 21-cm signal using 
the presently operating radio interferometers e.g. the 
GMRT\footnote{\url{http://www.gmrt.ncra.tifr.res.in}} \citep{paciga13}, 
LOFAR\footnote{\url{http://www.lofar.org}} \citep{haarlem13,yatawatta13},
the MWA\footnote{\url{http://www.mwatelescope.org}} \citep{jacobs16}, and
PAPER\footnote{\url{http://eor.berkeley.edu}} \citep{parsons14,jacobs14,ali15}.
The presently operating (first generation) radio interferometers are 
not sensitive enough to make tomographic images of the EoR 21-cm signal 
and can only make a statistical detection of the signal. Observing the 
EoR 21-cm signal is one of the major scientific goals of the upcoming 
radio telescopes e.g. SKA\footnote{\url{http://www.skatelescope.org}} 
\citep{mellema13,koopmans15} and HERA\footnote{\url{http://reionization.org}} 
\citep{deboer17}. These observations are very challenging due to the 
presence of foregrounds, system noise, and other calibration errors. 
Foregrounds are $\sim 4-5$ orders of magnitude stronger than the expected 
signal \citep{ali08,bernardi09,ghosh12,paciga13}, and modeling or 
removing them from the actual data is more complicated. However, in 
this work, we assume the idealistic scenario where foregrounds can be 
removed completely.

The upcoming SKA-Low will have 512 stations,\footnote{\label{ft:ska}\href{https://astronomers.skatelescope.org/wp-content/uploads/2016/09/SKA-TEL-SKO-0000422_02_SKA1_LowConfigurationCoordinates-1.pdf}{SKA1\_LowConfigurationCoordinates-1.pdf}} 
and each of them will be $\sim 35$~m in diameter. These stations will 
consist of several log-periodic dipole antennas. The telescope will also 
have $\sim 20\,{\rm deg}^2$ field of view, a compact core and 3 spiral 
arms which will extend up to $\sim 60$~km. SKA-Low will have enough 
sensitivity over a large range of frequencies (frequency band of 
$50 - 350$~MHz) to image the EoR 21-cm signal \citep{mellema15}. Unlike 
the CMB, we can map the large-scale structure~(LSS) of the universe in 
3D using the redshifted 21-cm signal, with the third dimension being 
frequency (or cosmic time or redshift). However, one has to be very 
careful while quantifying the EoR 21-cm signal as the mean, as well as 
other statistical properties of the signal change with varying frequency 
or redshift due to the light-cone~(LC) effect 
\citep{barkana06,datta12,laplante14,zawada14,mondal18}.

The LC effect breaks the statistical homogeneity (ergodicity) along 
the LoS direction. Moreover, the main assumption that goes into the 
estimation of the power spectrum $P(\kk)$ or equivalently into the 3D 
Fourier transform is that the signal is ergodic and periodic. As a 
consequence of this fundamental difference between the assumption for 
Fourier transform and the actual properties of the signal, the 
spherically averaged 3D power spectrum $P(k)$ fails to quantify the 
entire second-order statistics of the signal \citep{mondal18} and gives 
a rather biased estimation of the signal \citep{trott16}. This is 
particularly severe during the EoR when the $\xb$ changes rapidly 
as the reionization proceeds. This will also pose complications for 
broad bandwidth observations with SKA-Low \citep{mondal19}. The issue 
here is `how to quantify the statistics of the EoR 21-cm signal in the 
presence of the LC effect'. As a proper alternative to $P(k)$, we use 
the multi-frequency angular power spectrum~(MAPS) $\cl(\nu_1,\nu_2)$ 
\citep{datta07a,mondal18,mondal19}, which does not assume ergodicity and 
periodicity along the LoS. The only assumption is that the EoR 21-cm 
signal is statistically homogeneous and isotropic in different 
directions on the sky plane. The visibilities are the 
main observables in every radio-interferometric observations and 
the MAPS is directly associated with these visibility correlations. 
Therefore, it is relatively easy to estimate MAPS from the 
observations \citep{bharadwaj05,ali08,ghosh11}.

Several studies have been made to quantify the sensitivity for measuring 
the EoR 21-cm power spectrum with different instruments 
\citep{morales05,mcquinn06,zaroubi12,beard13,pober14,ewall16,shaw19}. 
These predictions were restricted to individually analyzing over small 
redshift (or equivalently frequency) intervals where they have worked 
with the 3D power spectrum $P(k)$. However, there is no such restriction 
for the MAPS, and we can, in principle, consider the entire bandwidth for 
the analysis. Here, we have made the SNR predictions for measuring the EoR 
21-cm MAPS using future observations with SKA-Low. We have presented 
our results mainly considering a scenario, the `Optimistic', where the 
observed MAPS is a sum of the EoR 21-cm MAPS and the system noise MAPS, 
ignoring any contribution from the foregrounds to the observed signal. 
However, we have also demonstrated the effects of foregrounds on the 
detection of the EoR 21-cm MAPS incorporating the foreground `wedge'. 
Note that we have used numerical simulations for computing the EoR 
21-cm MAPS in our analysis.

The paper is structured as follows. In Section~\ref{sec:sim}, we briefly 
describe the simulations used to generate the EoR 21-cm light-cones. 
Starting from the basic definition of the MAPS, we derive the 
expressions for the noise MAPS and MAPS error-covariance in 
Section~\ref{sec:maps}. In Section~\ref{sec:results}, we report the 
results i.e. the estimated MAPS, MAPS error-covariance and SNR assuming no foregrounds. Next, we discuss the impact of foregrounds on the prospects of detecting the 21-cm MAPS
in Section~\ref{sec:foreground}. Finally, in Section~\ref{sec:discuss}, we summarise our
results and conclude. Throughout the paper, we have used the values of
cosmological parameters $\Omega_{\rm m0}=0.27$, $\Omega_{\rm \Lambda0}=0.73$, 
$\Omega_{\rm b0}h^2=0.02156$, $h=0.7$, $\sigma_8=0.8$, and 
$n_{\rm s}=0.9619$. These values are consistent with the latest results 
from WMAP \citep{Komatsu2011} and Plank combined with other available 
constraints \citep{Planck2015,planck2016}.

%%%%%%%%%%%%%%%%%%%%%%%%%%%%%%%%%%%%%%%%%%%%%%%%%%%%%%%%%%%%%%%%%%%%%%%%%

\section{Simulating the E{\small o}R 21-{\small cm} signal}
\label{sec:sim}

\subsection{The Simulation}
\label{sec:sim1}
The density fields and halo catalogues are obtained from a 
high-resolution, large-volume $N$-body PRACE4LOFAR simulation 
\citep{giri2019}. This simulation was run using the 
\textsc{\small CubeP$^3$M} code \citep{Harnois2013} and followed 
$6912^3$ particles in a comoving 
$500\,h^{-1}{\rm Mpc}\approx 714\,{\rm Mpc}$ per side volume to enable 
reliable halo identification (with 25 particles or more) down to 
$10^9\, \rm M_\odot$. The reionization process is simulated using the 
\textsc{\small C$^2$-Ray} code \citep{methodpaper} on a $300^3$ grid 
with sources and density fields based on the $N$-body data following 
the method presented in \citet{Ilie07a} and \citet{Dixon:2016aa}. 
Specifically, for this work, we have used the data from the 
714Mpc\_g0.87\_gS\_300 reionization simulation following the 
notation of \citet{Dixon:2016aa}. We refer the reader to cited 
papers for details of the notation and setup, with only a brief 
summary provided here.

The density fields are calculated using SPH-like smoothing. The sources 
of ionization are associated with the resolved halos or high-mass
atomically cooling halos (HMACHs). These halos are complemented by 
a sub-grid model for the low-mass atomically-cooling halos (LMACHs)
$10^8<M_{\rm halo}<10^9$ \citep{Ahn15a}. Below this range, halos are 
assumed to not form stars. For a source with halo mass $M$ 
and lifetime $t_{\rm s}$, we assign ionizing photon emissivity according 
to
\be
\dot{N}_\gamma=g_\gamma\frac{M\Omega_{\rm b}}{\mu m_{\rm p}(10\,\rm Myr)\Omega_0} \,,
\ee
where the efficiency $g_{\gamma}$ combines the ionizing photon 
production efficiency of the stars per stellar atom, $N_{\rm i}$, the 
star formation efficiency, $f_*$, and the escape fraction, $f_{\rm esc}$:
\be
g_\gamma=f_*f_{\rm esc}N_{\rm i}\left(\frac{10 \;\mathrm{Myr}}{t_{\rm s}}\right).
\ee
\citep[e.g.][]{Haim03a,Ilie12a}. The high-mass sources ($M>10^9\,M_\odot$) 
are assumed unaffected by the radiative feedback and assigned an 
efficiency $g_{\gamma, \rm HMACH} = 0.87$. Prior to local reionization, 
the low-mass sources share the same efficiency as the high-mass sources. 
After the local ionization threshold exceeds 0.1, the low-mass sources 
have a mass-dependent efficiency
\be
g_{\gamma, \rm LMACH} \propto g_{\gamma, \rm HMACH} \times \left[\frac{M}{9\times10^8 \msun}-\frac{1}{9}\right] \,.
\label{eq:grad_eff}
\ee

%%%%%%%%%%%%%%%%%%%%%%%%%%%%%%%%%%%%%%%%%%%%%%%%%%%%%%%%%%%%%%%%%%%%%%%%%
\begin{figure}
\centering
\includegraphics[width=.45\textwidth]{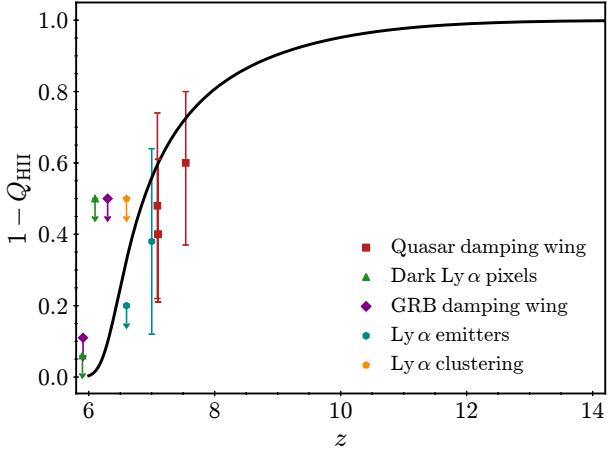}
\caption{The reionization history as a function of redshift obtained from our simulation. }
\label{fig:history}
\end{figure}
%%%%%%%%%%%%%%%%%%%%%%%%%%%%%%%%%%%%%%%%%%%%%%%%%%%%%%%%%%%%%%%%%%%%%%%%%

We have generated the coeval brightness temperature $(\delta \Tb)$ cubes 
at 125 different redshifts in the range $6\leq z <16$, and the resulting 
reionization history is shown in Figure~\ref{fig:history}. Our results are compared to observational inferences from Ly~$\alpha$ damping wings (squares; \protect\citealt{Greig2017,Davies2018,Greig2019}), dark Ly-$\rm \alpha$ forest pixels (triangles; \protect\citealt{McGreer2011, McGreer2015}), GRB damping wing absorption (diamonds; \protect\citealt{McQuinn2008, Chornock2013}), decline in Ly $\rm \alpha$ emitters (hexagons; \protect \citealt{Ota2008, Ouchi2010}), and Ly $\rm \alpha$ clustering (pentagons; \protect\citealt{Ouchi2010})
%%%%%%%%%%%%%%%%%%%%%%%%%%%%%%%%%%%%%%%%%%%%%%%%%%%%%%%%%%%%%%%%%%%%%%%%%
\subsection{Generating the light-cones}
\label{sec:LC}
We have generated our light cones following the formalism presented in 
\citet{datta14}, using the simulated coeval $\delta \Tb$ cubes described 
in Section~\ref{sec:sim1}. We have generated two light-cones: {\bf LC1} 
centered at $z_{\rm c}=7.09$ (frequency $\nu_{\rm c}=175.58$\,MHz), 
which corresponds to $\xb \approx 0.50$ and {\bf LC2} centered at 
$z_{\rm c}=8.04$ ($\nu_{\rm c}=157.08$\,MHz), which corresponds to 
$\xb \approx 0.75$. LC1 spans the redshift range $6.15 \la z \la 
8.25$, which corresponds to change in the mass-averaged \HI fraction 
$\xb$ (from end-to-end of the light cone, following the reionization 
history shown in Figure \ref{fig:history}) as $\Delta \xb \approx 
0.79-0.02=0.77$. Whereas, LC2 spans the range $6.92 \la z \la 9.40$, 
which corresponds to change in the $\xb$ as $\Delta \xb \approx 
0.90-0.42=0.48$. Note that the redshift ranges, channel widths and 
central frequencies assumed in the light-cones are only representative 
values and may change. We have chosen these to observe the behaviour at 
two different stages of reionization history.

%%%%%%%%%%%%%%%%%%%%%%%%%%%%%%%%%%%%%%%%%%%%%%%%%%%%%%%%%%%%%%%%%%%%%%%%%
\begin{figure}
\centering
\psfrag{Mpc}[c][c][1.5]{Mpc}
\includegraphics[width=.45\textwidth, angle=0]{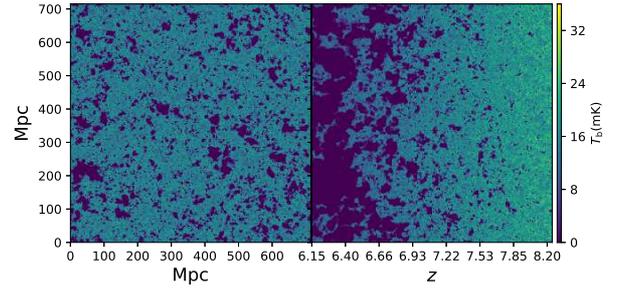}
\caption{This shows sections through the 3D 21-cm brightness
temperature maps for the coeval (left) and LC (right) simulations. 
The boxes are centered at redshift $7.09$ which corresponds to the 
comoving distance $r_{\rm c} = 8865.64\,{\rm Mpc}$ and $\xb \approx 0.50$.}
\label{fig:maps0.50}
\end{figure}

%%%%%%%%%%%%%%%%%%%%%%%%%%%%%%%%%%%%%%%%%%%%%%%%%%%%%%%%%%%%%%%%%%%%%%%%%

\begin{figure}
\centering
\includegraphics[width=.45\textwidth, angle=0]{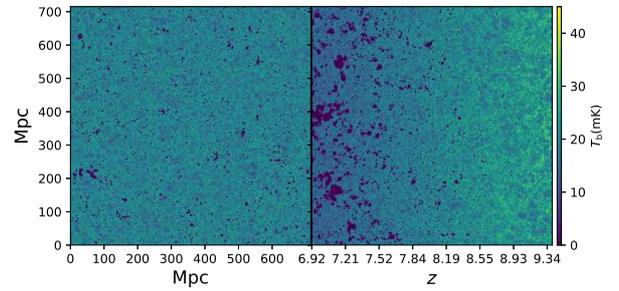}
\caption{Same as Figure~\ref{fig:maps0.50} centered at redshift $8.04$, 
which corresponds to the comoving distance $r_{\rm c} = 9162.06\,{\rm Mpc}$ 
and $\xb \approx 0.75$.}
\label{fig:maps0.75}
\end{figure}
%%%%%%%%%%%%%%%%%%%%%%%%%%%%%%%%%%%%%%%%%%%%%%%%%%%%%%%%%%%%%%%%%%%%%%%%%

The right panels of Figure~\ref{fig:maps0.50} and \ref{fig:maps0.75} 
show sections through the simulated LC 21-cm brightness temperature 
maps. As a comparison, the left panels of Figure~\ref{fig:maps0.50} and 
\ref{fig:maps0.75} show the sections through coeval simulations at 
$z=7.09$ and $z=8.04$, respectively. The lower redshifts on the left 
side of the LC simulations correspond to the later stages of the 
evolution as compared to the higher redshifts shown on the right side. 
The ionized regions appear smaller in the LC simulations as compared 
to their coeval companion at the right side (early stage). Whereas, the 
ionized regions appear larger in the LC simulations as compared to their 
coeval case at later stages (left side).

In this work, we assume the plane of the sky is flat. Under the 
flat-sky approximation, we map the brightness temperature fluctuations 
$\delta \Tb(x, y, z)$ from the Cartesian grid to a 3D rectangular grid 
in $(\thetavec, \nu)$ within our simulation box. We use 
$\theta_{\rm x}=x/r$, $\theta_{\rm y}=y/r$, and $\nu= z/r^{\prime}$. We 
also keep the angular extent the same at all frequency channels while 
performing this coordinate transformation.

%%%%%%%%%%%%%%%%%%%%%%%%%%%%%%%%%%%%%%%%%%%%%%%%%%%%%%%%%%%%%%%%%%%%%%%%%

\section{The Multifrequency Angular Power Spectrum}
\label{sec:maps}
The present study concerns the question: `How to quantify the statistics 
of the redshifted EoR \HI 21-cm signal $\delta \Tb (\n, \nu)$ when it 
is non-ergodic along the LoS (i.e. the signal varies significantly along 
the LoS)?' \citep{trott16,mondal18,mondal19}. In the case of the evolving 
statistical properties of the signal within the observed  volume, the 3D 
Fourier modes $\kk$ are not the correct choice of basis. Further, it 
assumes periodic boundary condition in all directions which is also not 
justified along the LoS. As a consequence, the power spectrum $P(\kk)$ is 
not optimal and gives a biased estimate of the true statistics 
\citep{trott16,mondal18}. To avoid this issue, the previous power 
spectrum measurements are restricted to individually 
analyzing small redshift intervals 
\citep{morales05,mcquinn06,zaroubi12,datta14,pober14,ewall16,shaw19}.

The above-mentioned properties of the signal necessitate the use 
the Multifrequency Angular Power Spectrum (MAPS) $\cl (\nu_1, \nu_2)$ 
which quantifies the entire second-order statistics of the EoR 21-cm 
signal \citep{mondal18}. It doesn't assume the signal to be 
statistically homogeneous along the LoS. One can decompose 
$\delta \Tb (\n, \nu)$ into spherical harmonics $Y_{\ell}^{\rm m}(\n)$ as
\begin{equation}
\delta T_{\rm b} (\n,\nu)=\sum_{\ell,m} a_{\ell {\rm m}} (\nu) \,
Y_{\ell}^{\rm m}(\n)\, ,
\label{eq:a1}
\end{equation}
and define the MAPS using 
\begin{equation}
\cl(\nu_1, \nu_2) = \big\langle a_{\ell {\rm m}} (\nu_1)\, a^*_{\ell 
{\rm m}} (\nu_2) \big\rangle\, .
\label{eq:cl}
\end{equation}
The only assumption that goes into this definition is that the EoR 21-cm 
signal is statistically homogeneous and isotropic in different directions 
on the sky plane.

In this study, we have chosen to work in the flat-sky approximation, 
where the redshifted 21-cm brightness temperature fluctuations can be 
expressed as $\delta \Tb (\thetavec, \nu)$. Here $\thetavec$ denotes a 
2D vector on the plane of the sky. Instead of $\delta \Tb (\thetavec, \nu)$, 
we use its 2D Fourier transform $\TTb(\U,\nu)$, where $\U$ is the Fourier 
conjugate of $\thetavec$ described in the previous section. 
$\TTb(\U,\nu)$ is the primary observable measured in radio 
interferometric observations. Under the flat-sky approximation, we 
redefine the MAPS (eq.~\ref{eq:cl}) as
\begin{equation}
\cl(\nu_1,\nu_2) \equiv {\mathcal C}_{2\upi{\rm U}}(\nu_1,\nu_2) =
\Omega^{-1}\,\big\langle \TTb(\U,\nu_1)\,
\TTb(-\U,\nu_2)\big\rangle\,,
\label{eq:cl_flat}
\end{equation}
where $\Omega$ is the solid angle subtended by the transverse extent of 
the observation (or simulation) at the location of the observer and
$\ell=2\upi{\rm U}$ is the corresponding angular multipole. The above 
definition of $\cl(\nu_1,\nu_2)$ does not assume statistical 
homogeneity and periodicity along the LoS. However, note that if one 
imposes statistical homogeneity along the LoS, the MAPS $\cl(\nu_1,\nu_2)$ 
is expected to depend only on the frequency separation 
$\Delta \nu = |\nu_1 - \nu_2|$, i.e. $\cl(\nu_1,\nu_2) \equiv \cl(\Delta \nu)$. 

%The visibilities $\TTb(\U,\nu)$, which are 2D Fourier transform of $\delta \Tb(\thetavec,\nu)$, are the observables at baselines $\U=\dd/\lambda_{\rm i}$ in every radio-interferometric observations. Here $\dd$ is the projected antenna separation on the sky plane perpendicular to the LoS and $\lambda_{\rm i}$ is the wavelength corresponds to the frequency $\nu_{\rm i}$.

%%%%%%%%%%%%%%%%%%%%%%%%%%%%%%%%%%%%%%%%%%%%%%%%%%%%%%%%%%%%%%%%%%%%%%%%%
\setlength{\unitlength}{1cm}
\begin{figure*}
\centering
\psfrag{nu2}[c][c][1.2]{$\nu_2 - \nu_{\rm c}$ (MHz)}
\psfrag{nu1}[c][c][1.2]{$\nu_1 - \nu_{\rm c}$ (MHz)}
\psfrag{cl}[c][c][1.2]{$\Phi^2(\nu_1, \nu_2)~({\rm mK})^2$}

\includegraphics[width=.95\textwidth, angle=0]{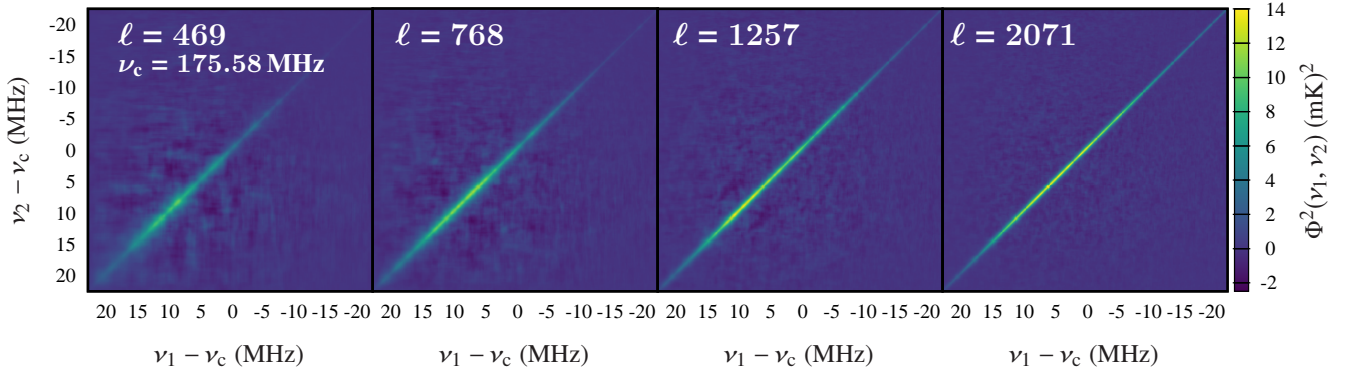}
\put(-15.60,4.2){{\Large \textcolor{white}{\bm{$\ell = 469$}}}}
\put(-11.95,4.2){{\Large \textcolor{white}{\bm{$\ell = 768$}}}}
\put(-8.30,4.2){{\Large \textcolor{white}{\bm{$\ell = 1257$}}}}
\put(-4.65,4.2){{\Large \textcolor{white}{\bm{$\ell = 2071$}}}}
\put(-15.60,3.85){{\large \textcolor{white}{\bm{$\nu_{\rm c}=175.58\,{\rm MHz}$}}}}
\caption{This shows the MAPS $\Phi^2(\nu_1, \nu_2)$ at $\ell = 469$, $768$, $1257$ and $2071$ (from left to right respectively) for the LC1 at $\nu_{\rm c}=175.58\,$MHz.}
\label{fig:cl0.50}
\end{figure*}

%%%%%%%%%%%%%%%%%%%%%%%%%%%%%%%%%%%%%%%%%%%%%%%%%%%%%%%%%%%%%%%%%%%%%%%%%
\setlength{\unitlength}{1cm}
\begin{figure*}
\centering
\psfrag{nu2}[c][c][1.2]{$\nu_2 - \nu_{\rm c}$ (MHz)}
\psfrag{nu1}[c][c][1.2]{$\nu_1 - \nu_{\rm c}$ (MHz)}
\psfrag{cl}[c][c][1.1]{$\Phi^2(\nu_1, \nu_2)~({\rm mK})^2$}

\includegraphics[width=0.95\textwidth, angle=0]{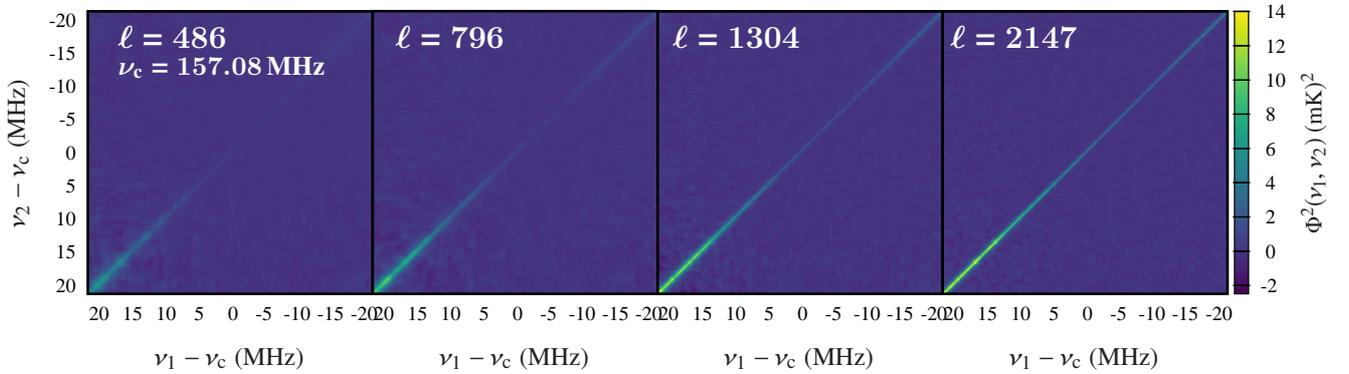}
\put(-15.60,4.2){{\Large \textcolor{white}{\bm{$\ell = 486$}}}}
\put(-11.95,4.2){{\Large \textcolor{white}{\bm{$\ell = 796$}}}}
\put(-8.30,4.2){{\Large \textcolor{white}{\bm{$\ell = 1304$}}}}
\put(-4.65,4.2){{\Large \textcolor{white}{\bm{$\ell = 2147$}}}}
\put(-15.60,3.85){{\large \textcolor{white}{\bm{$\nu_{\rm c}=157.08\,{\rm MHz}$}}}}

\caption{Same as Figure~\ref{fig:cl0.50} at $\ell = 486$, $796$, $1304$ and 
$2147$ (from left to right respectively) for the LC2 at $\nu_{\rm c}=157.08\,{\rm MHz}$.}
\label{fig:cl0.75}
\end{figure*}
%%%%%%%%%%%%%%%%%%%%%%%%%%%%%%%%%%%%%%%%%%%%%%%%%%%%%%%%%%%%%%%%%%%%%%%%%

The $\ell$ range of each light cone is divided into $10$ equally spaced logarithmic bins in our analysis, and each bin is tagged by the bin-averaged value of $\ell$ \textit{i.e.} $\ell_i$ for the $i$-th bin. Note that the average value $\ell_i$ varies from LC1 to LC2. In this work, we focus mainly on the intermediate $\ell$ bins as the detection of the signal will be difficult at large scales ($\ell$ $\la 250$) due to the cosmic variance and at small scales ( $\ell$ $\ga 3500$) due to the presence of large system noise. For the LC1, we have shown our results at four different $\ell$ bins that are $\ell = 469,\,768,\,1257$ and $2071$. These $\ell$ values roughly correspond to comoving scales $119\,{\rm Mpc}$, $72\,{\rm Mpc}$, $44\,{\rm Mpc}$ and $26\,{\rm Mpc}$ respectively at the central frequency $\nu_{\rm c}=175.58$\,MHz of LC1. For the LC2, we have shown the results at values of $\ell = 486,\,796,\,1304$ and $2147$, which roughly correspond to almost the same comoving scales as for the LC1, at the central frequency $\nu_{\rm c}=157.08$\,MHz.
Figure~\ref{fig:cl0.50} shows the scaled MAPS 
$\Phi^2(\nu_1, \nu_2)=[\ell(\ell+1)\cl(\nu_1, \nu_2)/2\upi]$ at four 
aforementioned values of $\ell$ for the LC1 simulation, which is 
centered at a redshift having $\xb \approx 0.50$. Figure~\ref{fig:cl0.75} shows the same 
for the LC2 simulation, which is centered at a redshift with $\xb \approx 0.75$. 
We see that the MAPS peaks when $\nu_1=\nu_2$, i.e. along 
the diagonal line. The diagonal $\cl(\nu, \nu)$ evolves considerably with 
the observed frequency $\nu$. This is a direct consequence of the fact 
that the signal is non-ergodic along the frequency axis. We also find that 
the MAPS rapidly falls as the frequency separation $\mid \nu_1 - \nu_2 \mid$
increases and oscillates around zero for the larger frequency separation. 
Unlike the 3D power spectrum $P(\kk)$, which captures only the information 
regarding the ergodic and periodic part of the signal, the MAPS 
$\cl(\nu_1,\nu_2)$ contains the full information regarding the two-point 
statistics of the signal \citep{mondal18}.  One can, in principle, use 
the entire information contained in $\cl(\nu_1,\nu_2)$, i.e. all the 
diagonal and off-diagonal elements to better constrain the EoR. However, 
we focus mostly on the diagonal terms $\cl(\nu, \nu)$. It will be 
difficult to detect the off-diagonal $\cl$s, except for the small 
frequency separation $\mid \nu_1 - \nu_2 \mid \sim 1$\,MHz, due to poor signal 
to noise ratio. We refer the readers to Section~\ref{sec:results} for a
detailed discussion on the detectability of the MAPS. 

%%%%%%%%%%%%%%%%%%%%%%%%%%%%%%%%%%%%%%%%%%%%%%%%%%%%%%%%%%%%%%%%%%%%%%%%%
\begin{figure*}
\centering
\psfrag{l=469}[c][c][1.]{$\ell=469$}
\psfrag{l=768}[c][c][1.]{$\ell=768$}
\psfrag{l=1257}[c][c][1.]{$\ell=1257$}
\psfrag{l=2071}[c][c][1.]{$\ell=2071$}

\psfrag{l=486}[c][c][1.]{$\ell=486$}
\psfrag{l=796}[c][c][1.]{$\ell=796$}
\psfrag{l=1304}[c][c][1.]{$\ell=1304$}
\psfrag{l=2147}[c][c][1.]{$\ell=2147$}

\psfrag{xh1}[c][c][1.2]{$\xb$}
\psfrag{nu}[c][c][1.2]{$\nu$ (MHz)}
\psfrag{cl}[c][c][1.1]{$\Phi^2(\nu, \nu)~({\rm mK})^2$}

\includegraphics[width=0.99\textwidth, angle=0]{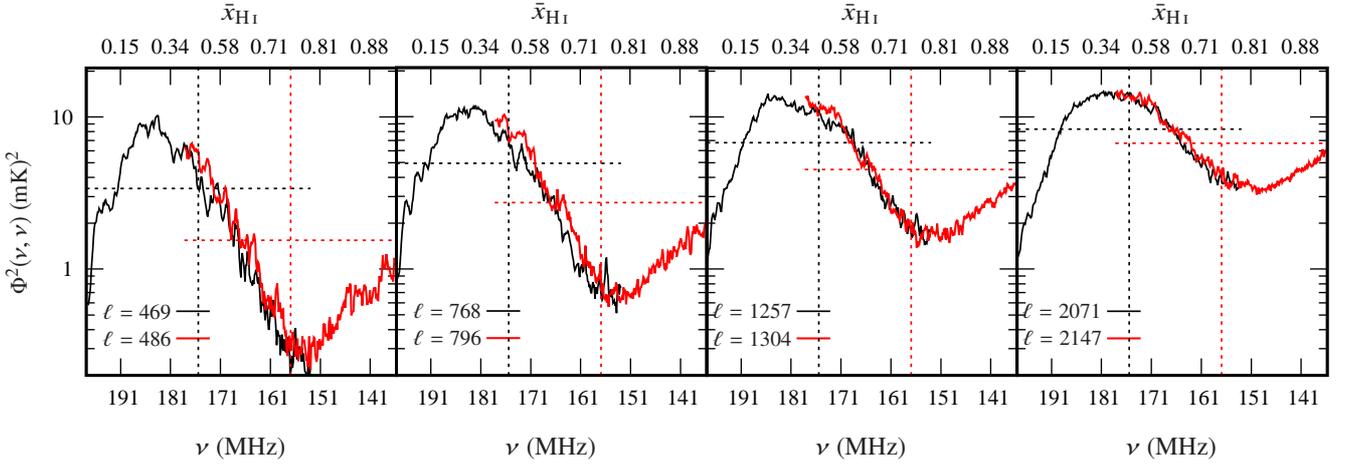}

\caption{This shows the diagonal components of the scale-independent MAPS $\Phi^2(\nu, \nu)$ for LC1 (black) and LC2 (red). The LC1 and LC2 are respectively centered at frequency $175.58\,$MHz and $157.08\,$MHz (vertical dashed lines). We also show the ergodic component (mean) of MAPS $[\ell(\ell+1)\cl^{\rm E}(\nu, \nu)/2\upi]$ (horizontal dashed lines). The $\xb$ values corresponding to the frequencies are shown in the top $x$-axis.}
\label{fig:cl_0.50_0.75}
\end{figure*}
%%%%%%%%%%%%%%%%%%%%%%%%%%%%%%%%%%%%%%%%%%%%%%%%%%%%%%%%%%%%%%%%%%%%%%%%%

Figure~\ref{fig:cl_0.50_0.75} shows the diagonal components of the scaled 
MAPS $\Phi^2(\nu, \nu)$ as a function of $\nu$ for $\ell$ values 
considered above for both simulations LC1 (black) and LC2 (red). It also
shows the ergodic part  of the signal $\Phi^2(\nu_c, \nu_c)$ that is 
calculated at the central frequency $\nu_c$, which is different for the 
LC1 and LC2  simulation. The 3D power 
spectrum $P(\kk)$ misses the part that is deviated from these 
horizontal dashed lines. We further see in Figure~\ref{fig:cl_0.50_0.75} 
that $\Phi^2(\nu, \nu)$ peaks around a frequency corresponding 
to the global neutral fraction $\xb \approx 0.35$ for both simulations. 
This is due to the  presence of a significant number of large ionized 
bubbles at that stage of the EoR. The power spectrum at higher frequency
decreases due to the rapid decline of the neutral fraction $\xb$. The
characteristic size of ionized bubbles decreases at lower frequencies, 
which causes the power spectrum to decrease. Similar results have been 
found in earlier studies \citep{mcquinn07,lidz08,choudhury09b,mesinger11}. 
We also notice  that there is a `dip' in the power spectrum 
$\Phi^2(\nu, \nu)$ around a frequency corresponding to the global 
neutral fraction $\xb \approx 0.8$ for all $\ell$ modes for both 
simulations. During the early stages of reionization, the high-density 
regions get ionized first, and as a consequence, the large-scale power 
decreases. This is reflected by the drop in the power across the four 
$\ell$ panels when the neutral fraction is large. Later, as the 
reionization progresses further, the creation and growth of the ionized 
regions increase the power spectrum which peaks around 
$\xb \approx 0.35$.  \citet{datta14} have investigated the impact 
of the LC effect considering a similar reionization model and find a 
similar dip around $\xb \sim 0.8$. The  frequencies at which the minimum 
and maximum occur may  change for different $\ell$ values. However, we do
not see any significant change in the locations of the maxima and minima 
for the $\ell$ modes we consider.

%%%%%%%%%%%%%%%%%%%%%%%%%%%%%%%%%%%%%%%%%%%%%%%%%%%%%%%%%%%%%%%%%%%%%%%%%
\begin{figure}
\centering
\psfrag{nu=185}[c][c][1.]{$\nu=185$\,MHz~}
\psfrag{nu=170}[c][c][1.]{$\nu=170$\,MHz~}
\psfrag{nu=155}[c][c][1.]{$\nu=155$\,MHz~}
\psfrag{nu=140}[c][c][1.]{$\nu=140$\,MHz~}

\psfrag{l}[c][c][1.2]{$\ell$}
\psfrag{cl}[c][c][1.1]{$\ell(\ell+1)\cl/2\upi~~({\rm mK})^2$}

\includegraphics[width=0.5\textwidth, angle=0]{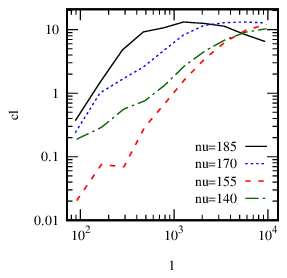}

\caption{This shows the angular power spectrum of the EoR 21-cm brightness temperature fluctuations as a function of $\ell$ at four different frequency $\nu=185$\,MHz (LC1), 170\,MHz (LC1), 155\,MHz (LC2) and 140\,MHz (LC2) for a particular case of MAPS where $\nu_1=\nu_2$.}
\label{fig:cl_l}
\end{figure}
%%%%%%%%%%%%%%%%%%%%%%%%%%%%%%%%%%%%%%%%%%%%%%%%%%%%%%%%%%%%%%%%%%%%%%%%%

The Figure~\ref{fig:cl_l} shows the angular power spectrum of the EoR 21-cm brightness temperature fluctuations as a function of $\ell$ at four different frequency $\nu=185$\,MHz (LC1), 170\,MHz (LC1), 155\,MHz (LC2) and 140\,MHz (LC2). Note that the angular power spectrum shown in this plot is a spacial case of MAPS where $\nu_1=\nu_2$. The basic assumption of our model is that the hydrogen traces the underlying dark matter distribution. As a result of this, the shape of the 21-cm angular power spectra is roughly the same as the dark matter angular power spectrum at the start of reionization ($\nu \ga 140$\,MHz). As discussed above (Figure~\ref{fig:cl_0.50_0.75}), during the early stages of reionization, the high-density regions get ionized first in the inside-out scenario, and as a consequence, the power spectrum drops at $\nu=155$\,MHz. As reionization progresses, the creation and growth of the ionized bubbles increase the power at $\nu=170$\,MHz. The contrast of the brightness temperature fluctuation field peaks on large scales at $\nu \approx 185$\,MHz which corresponds to $\xb \approx 0.35$ in our fiducial reionization model. This further raises the power of signal at small $\ell$ (large length scales). However growth of the ionized regions reduces contrast of the signal at small length scales showing drop in power at the corresponding $\ell$ values.

%%%%%%%%%%%%%%%%%%%%%%%%%%%%%%%%%%%%%%%%%%%%%%%%%%%%%%%%%%%%%%%%%%%%%%%%%

\section{Observational considerations}
\label{sec:obs}
We now consider  observations with a radio-interferometric array where 
the fundamental quantity is the visibility that is measured by each pair 
of antennas in the array. Considering any  particular pair with 
$\dd_{\rm n}$ being the antenna separation projected  on the plane 
perpendicular to the LoS,  the visibility measured at frequency 
$\nu_{\rm i}$ and  baseline  $\U_{\rm n}=\dd_{\rm n}/\lambda_{\rm i}$ 
provides a direct estimate of $\TTb(\U_{\rm n},\nu_{\rm i})$ at the 
Fourier mode $\U_{\rm n}$. Taking into account $\TTb^{\rm N}(\U,\nu)$ 
the system noise contribution which is inherent in any 
radio-interferometric observation, the measured visibility 
actually provides us with 
$\TTb^{\rm t}(\U_{\rm n},\nu_{\rm i})=\TTb(\U_{\rm n},\nu_{\rm i})+
\TTb^{\rm N}(\U_{\rm n},\nu_{\rm i})$, where we have assumed that the 
foregrounds have been completely removed and there are no calibration 
errors. The system  noise at different baselines and frequency channels is 
uncorrelated. Using this in eq.~(\ref{eq:cl_flat}) for the MAPS, we obtain
\begin{equation}
\cl^t(\nu_1,\nu_2)=\cl(\nu_1,\nu_2) + 
\delta^{\rm K}_{\nu_1\nu_2}\,\cl^{\rm N}(\nu_1,\nu_2)\,,
\label{eq:clsum}
\end{equation}  
which can be estimated from the observed visibilities. Following the 
prescription in \citet{3DTGE}, it is possible to avoid noise bias 
$\cl^{\rm N}(\nu_1,\nu_2)$ and obtain an unbiased estimate of 
$\cl(\nu_1,\nu_2)$ from the measured visibilities. However, the 
noise contributions still persist in the error estimates and this 
cannot be avoided. In this work, we compute the error variance to 
predict the signal-to-noise ratio~(SNR) of measuring MAPS using the 
upcoming SKA-Low. This also involves the estimation of system noise for 
which we use the telescope specifications of SKA-Low taken from the 
current proposed configuration document\textsuperscript{\ref{ft:ska}}. 
Some important specifications\footnote{The specifications assumed here 
may change in the final implementation of the telescope.} that have 
been used in the computation of $\cl^{\rm N}(\nu,\nu)$ are tabulated 
in Table~\ref{table:ska} . 

%%%%%%%%%%%%%%%%%%%%%%%%%%%%%%%%%%%%%%%%%%%%%%%%%%%%%%%%%%%%%%%%%%%%%%%%%
\begin{table}
\centering
\caption{This tabulates the telescope specifications for the current proposed 
configuration of SKA-Low}
\label{table:ska}
\begin{tabular}{cc}
\hline
\hline
\,Parameters\, & \,Values\,\rule{0pt}{2.6ex}\rule[-1.2ex]{0pt}{0pt}\\
\hline
Number of stations & $512$ \rule{0pt}{2.6ex}\rule[-1.2ex]{0pt}{0pt}\\
Diameter of each station $(D)$ & 35\,meters \rule{0pt}{2.6ex}\rule[-1.2ex]{0pt}{0pt}\\ 
Operation frequency range & $50-350$\,MHz \rule{0pt}{2.6ex}\rule[-1.2ex]{0pt}{0pt}\\
Receiver temperature ($T_{\rm rec}$) & $100$\,K \rule{0pt}{2.6ex}\rule[-1.2ex]{0pt}{0pt}\\
Maximum baseline separation & $\sim 19\,{\rm km}$ \rule{0pt}{2.6ex}\rule[-1.2ex]{0pt}{0pt}\\
\hline
\end{tabular}
\end{table}
%%%%%%%%%%%%%%%%%%%%%%%%%%%%%%%%%%%%%%%%%%%%%%%%%%%%%%%%%%%%%%%%%%%%%%%%%

\begin{figure}
\centering
\includegraphics[width=0.45\textwidth, angle=0]{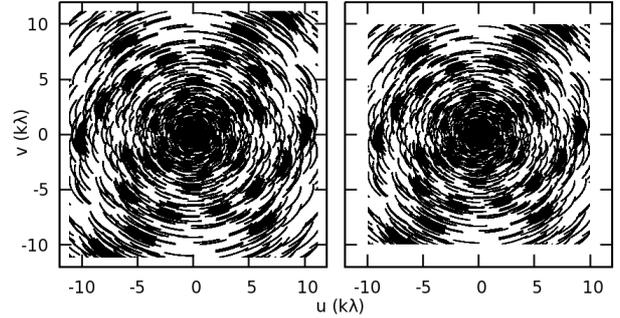} 
\caption{The SKA-Low $uv$ coverage with phase centre at RA$=13\,{\rm hr}~2\,{\rm m}~31.5\,{\rm s}$ and DEC$=-26{\degr}~49^{\prime}~29^{\prime \prime}$ for a total observation time of $2\,$hrs. $u$ and $v$ are projected antenna separation (for $\mid \dd \mid < 19$\,km) in the unit of k$\lambda$ at the central frequencies $\nu_{\rm c}=175.58$\,MHz (left) and $\nu_{\rm c}=157.08$\,MHz (right).}
\label{fig:uv}
\end{figure}

We consider  observations tracking a field at declination 
DEC$=-30{\degr}$ for 8\,hrs/night with 60\,sec integration time 
following the formalism adopted by \citet{shaw19}. We restrict our 
analysis to the baselines $\U$ corresponding to the antenna 
separations $\mid \dd \mid < 19$\,km, as the baseline distribution 
falls off rapidly at larger $\mid \dd \mid$ values. 
Figure~\ref{fig:uv} shows the  simulated SKA-Low baseline $\U$ 
distribution ($uv$ coverage) at the two different central frequencies,
corresponding to LC1 and LC2 respectively. The signals at two different 
baselines $\U$ separated by $<D/\lambda_{\rm i}$ are correlated due to 
the overlap of the antenna beam pattern \citep{bharadwaj2003,bharadwaj05}. 
We grid the baselines $\U_{\rm m}$ with a grid of size 
$\Delta U_{\rm x}=\Delta U_{\rm y}=D/\lambda_{\rm i}$ and count the 
number of measurements $\tau(\U_{\rm g})$ that lie within a pixel 
centered at any  grid point $\U_{\rm g}$. 

%%%%%%%%%%%%%%%%%%%%%%%%%%%%%%%%%%%%%%%%%%%%%%%%%%%%%%%%%%%%%%%%%%%%%%%%%
%%%%%%%%%%%%%%%%%%%%%%%%%%%%%%%%%%%%%%%%%%%%%%%%%%%%%%%%%%%%%%%%%%%%%%%%%

We estimate the noise MAPS at the grid point $\U_{\rm g}$ following 
the calculation presented in \citet{white99, zaldarriaga04} and 
\citet{shaw19} as
\begin{align}
\cc_{\ell_{\rm g}}^{\rm N}(\nu,\nu) &= \frac{T^2_{\rm sys} \, \lambda^4}{N_{\rm p}\,N_{\rm t} \,
\Delta t\,\Delta \nu\,a^2\,\tau(\U_{\rm g})}\times \frac{1}{\int d\U^{\prime}\,
\mid \tilde{A}(\U - \U^{\prime})\mid^2} \nonumber \\
&= \frac{8\,{\rm hrs}}{t_{\rm obs}} \times \frac{{\mathcal C}^{\rm 0}(\nu)}{\tau(\U_{\rm g})}\,.
\label{eq:cln}
\end{align}
Here the system temperature $T_{\rm sys}$ is a sum of the sky temperature
$T_{\rm sky}=60\lambda^{2.55}$\,K \citep{fixsen11} and the receiver 
temperature $T_{\rm rec}$. $N_{\rm p}$ is the number of polarizations, 
$N_{\rm t}$ is the number of observed nights, $\Delta t$ is the 
integration time, $a$ is the area of individual antenna in the array and 
$\tilde{A}(\U)$ is the Fourier transform of the primary beam of a 
station $A(\thetavec)$, which is approximated with a Gaussian 
$e^{-(\theta/\theta_{\rm o})^2}$ \citep{Samir_2014,shaw19}. We express 
the total observation time using $t_{\rm obs}= 8\,{\rm hrs}\times 
N_{\rm t}$, and this notation is used in the rest of the paper. 

%%%%%%%%%%%%%%%%%%%%%%%%%%%%%%%%%%%%%%%%%%%%%%%%%%%%%%%%%%%%%%%%%%%%%%%%%

\subsection{The binned weighted MAPS estimator}
\label{sec:estimator}
The simulated observations under consideration have $\sim 300\times 
300$ grid points on the $\U$ plane and $313$ frequency channels. This 
comes out to $\sim 14$ Million independent measurements of the MAPS which 
is computationally very expensive to deal with. Another problem is that 
the measurements at every individual grid point $\U_{\rm g}$ will be 
very noisy. To tackle these issues, we bin the U space. We, however, 
lose the information at individual $\U_{\rm g}$ modes. This not only 
solves the computation problem but also increases the SNR of measurement 
within a bin. We use the binned weighted MAPS estimator 
$\hat{\cl}_{\rm i}^{\rm t}(\nu_1,\nu_2)$, which is the sum of the 
weighted brightness temperature fluctuation correlations between various 
grids within the bin. Exploiting the symmetry 
$\hat{\cl}_{\rm i}^{\rm t}(\nu_1,\nu_2)=\hat{\cl}_{\rm i}^{\rm t}(\nu_2,\nu_1)$, 
the estimator $\hat{\cl}^{\rm t}(\nu_1,\nu_2)$ for the $i$-th bin is 
written as
\begin{align}
\hat{\cl}_{\rm i}^{\rm t}(\nu_1,\nu_2)=\frac{1}{2\Omega} \sum_{\U_{\rm g_i}} 
\hat{w}(&\U_{\rm g_i},\nu_1)\, \hat{w}(\U_{\rm g_i},\nu_2)\times \nonumber \\
&[\TTb^{\rm t}(\U_{\rm g_i},\nu_1)\, \TTb^{\rm t}(-\U_{\rm g_i},\nu_2) \nonumber \\
&+\TTb^{\rm t}(\U_{\rm g_i},\nu_2)\, \TTb^{\rm t}(-\U_{\rm g_i},\nu_1)]\,,
\label{eq:est1}
\end{align}
where the sum $\sum_{\U_{\rm g_i}}$ is over the $\U_{\rm g}$ grids within
the $i$-th bin and $\hat{w}(\U_{\rm g},\nu)$ is the weight associated 
with the grid $\U_{\rm g}$ at frequency $\nu$. Here the angular 
multipole $\ell_{\rm i}=2\upi U_{\rm i}$ (or $U_{\rm i}$) is the weighted 
average of all $\U_{\rm g}$ in the $i$-th bin. We have used equally 
spaced logarithmic binning, and the bins here are semi-annuli of the 
width $\Delta U_{\rm i} \propto U_{\rm i}$ (restricted to one half of 
the $\U$ plane as the signal is real, i.e. $\TTb^{\rm t*}(\U_{\rm g_i},\nu_1)=\TTb^{\rm t}(-\U_{\rm g_i},\nu_1)$).

The ensemble average of the estimator gives the bin-averaged MAPS
\begin{align}
\langle \hat{\cl}_{\rm i}^{\rm t}(\nu_1,\nu_2) \rangle 
&\equiv \bar{\cl}_{\rm i}^{\rm t}(\nu_1,\nu_2)  \nonumber \\
&= \bar{\cl}_{\rm i}(\nu_1,\nu_2) + 
\delta^{\rm K}_{\nu_1\nu_2}\, \bar{\cl}_{\rm i}^{\rm N}(\nu_1,\nu_2)\,.
\end{align}
As  mentioned earlier, it is possible to avoid the noise bias 
$\bar{\cl}^{\rm N}(\nu,\nu)$ \citep{3DTGE} by subtracting out the
contribution of the self-correlation of visibility from itself. 
This also leads to a loss of a part of the signal. However, 
this loss is extremely small $(<0.01 \%)$ for long observations 
($t_{\rm obs}\sim 100\,{\rm hrs}$ or larger) with $16 \, {\rm s}$ 
integration time. It is therefore quite well justified to assume 
that we can obtain  an unbiased estimate of $\bar{\cl}(\nu_1,\nu_2)$. 
In the subsequent analysis, we also do not consider any change in the 
weights along the frequency direction and express eq.~\ref{eq:est1} as
\begin{align}
\hat{\cl}_{\rm i}^{\rm t}(\nu_1,\nu_2) = \frac{1}{2\Omega} \sum_{\U_{\rm g_i}} 
& w(\U_{\rm g_i})\, [\TTb^{\rm t}(\U_{\rm g_i},\nu_1)\, 
\TTb^{\rm t}(-\U_{\rm g_i},\nu_2) \nonumber \\
&+\TTb^{\rm t}(\U_{\rm g_i},\nu_2)\, \TTb^{\rm t}(-\U_{\rm g_i},\nu_1)]\,.
\label{eq:est2}
\end{align}
The  weights $w(\U_{\rm g_i})$ are normalized such that 
$\sum_{\U_{\rm g_i}} w_{\rm g_i} = 1$ where the sum runs over each 
grid point within a particular $U$ bin. As discussed later, the weights 
are selected in order to maximize the SNR of 
$\bar{\cl}_{\rm i}(\nu_1,\nu_2)$ for each bin. This takes into account 
that the baselines $\U_{\rm m}$ do not uniformly sample the different 
grid points $\U_g$, and consequently the ratio 
$\cc_{\ell_{\rm g_i}}(\nu_1,\nu_2)/\cc_{\ell_{\rm g_i}}^{\rm 
N}(\nu_1,\nu_2)$ varies  across the different grid points within a bin.

%%%%%%%%%%%%%%%%%%%%%%%%%%%%%%%%%%%%%%%%%%%%%%%%%%%%%%%%%%%%%%%%%%%%%%%%%

\subsection{The error estimates}
\label{sec:error}
The EoR 21-cm signal is a highly non-Gaussian field (see e.g. \citealt{bharadwaj05a,mondal15,yoshiura15,majumdar18}). The non-Gaussian effects will play a significant role in the error estimates for the EoR 21-cm MAPS. The Gaussian components in observed visibilities $\TTb(\U,\nu)$ are independent at different baselines. It is the non-Gaussian components which are correlated and give rise to non-zero higher order statistics such as bispectrum, trispectrum etc. The cosmic variance of the MAPS will get additional contributions from the non-zero trispectra (see eq. 2 of \citealt{mondal17}) which is essentially the Fourier conjugate of the four-point correlation function (see also e.g. eq. 2 of \citealt{adhikari2019}). The angular trispectra will also introduce correlations between the errors in MAPS at different $\ell$ modes. \citet{mondal16} and \citet{mondal17} have quantitatively demonstrated the impact of non-Gaussianity in the context of estimating the EoR 21-cm 3D power spectrum cosmic variance. Recently \citet{shaw19} have shown that, in presence of the Gaussian system noise and foreground contamination, the contribution of trispectrum to the error variance is significant within a limited range of $k$ modes, and mostly during the later stages of reionization ($z\lesssim 10$). Considering the SKA-Low observations, we can ignore the contribution of the non-Gaussianity of the signal in the observed MAPS error covariance. Furthermore, estimating the angular trispectrum of the light-cone signal is computationally challenging. Hence for simplicity, we do not consider the non-Gaussian nature of the EoR 21-cm signal in our calculations and assume the error estimate of the MAPS is completely determined by that of the Gaussian random field predictions. Following the calculation presented in  
Appendix~\ref{a1}, we write the MAPS error covariance as 
%\citet{datta07a} show that the EoR signal itself decorrelates when $\Delta %\nu \gtrsim 0.1~{\rm MHz}$ for $\ell \sim 10^3$. Thus we expect the %correlations and also the trispectrum to decay rapidly with $\Delta \nu$. 
\begin{align}
\cov_{12,34}&=\langle[\delta \cc_{\ell_{\rm i}}^{\rm t}(\nu_1,\nu_2)]
[\delta \cc_{\ell_{\rm i}}^{\rm t}(\nu_3,\nu_4)]\rangle \nonumber \\
&=\frac{1}{2} \sum_{\U_{\rm g_i}} w_{\rm g_i}^{2}\, 
[\cc_{\ell_{\rm g_i}}^{\rm t}(\nu_1,\nu_3)\cc_{\ell_{\rm g_i}}^{\rm t}(\nu_2,\nu_4) \nonumber \\
&\tab +\cc_{\ell_{\rm g_i}}^{\rm t}(\nu_1,\nu_4)\cc_{\ell_{\rm g_i}}^{\rm t}(\nu_2,\nu_3)]\,,
\label{eq:cov}
\end{align}
where the sum is over all the $\U_{g_i}$ grids within the $i$-th bin and 
$w_{\rm g_i}\equiv w(\U_{\rm g_i})$. The variance in the measured 
$\bar{\cl}_{\rm i}^{\rm t}(\nu_1,\nu_2)$ is thus given by 
\begin{align}
\cov_{12,12}&=[\bm{\sigma}_{12}^{\ell_{\rm i}}]^2=\langle [\delta 
\cc_{\ell_{\rm i}}^{\rm t}(\nu_1,\nu_2)]^2 \rangle \nonumber \\
&= \frac{1}{2}\sum_{\U_{\rm g_i}} w^2_{\rm g_i} 
[\cc_{\ell_{\rm g_i}}^{\rm t}(\nu_1,\nu_1)\cc_{\ell_{\rm g_i}}^{\rm t}(\nu_2,\nu_2)
+\{\cc_{\ell_{\rm g_i}}^{\rm t}(\nu_1,\nu_2)\}^2] \nonumber \\
&= \frac{1}{2} \sum_{\U_{\rm g_i}} w^2_{\rm g_i}\,[\{\cc_{\ell_{\rm g_i}}(\nu_1,\nu_1)
+ \cc_{\ell_{\rm g_i}}^{\rm N}(\nu_1,\nu_1)\} \nonumber \\ 
&\tab \times \{\cc_{\ell_{\rm g_i}}(\nu_2,\nu_2) + 
\cc_{\ell_{\rm g_i}}^{\rm N}(\nu_2,\nu_2)\} + \nonumber \\
&\tab \{\cc_{\ell_{\rm g_i}}(\nu_1,\nu_2) + \delta^{\rm K}_{\nu_1\nu_2}\, \cc_{\ell_{\rm g_i}}^{\rm N}(\nu_1,\nu_2)\}^2]\,,
\label{eq:err1}
\end{align}
where the sum is over the grids points $\U_{\rm g_i}$ within the $i$-th 
bin. We know that the MAPS signal peaks along the diagonal elements 
$\nu_1=\nu_2$, where the error variance (using eq.~\ref{eq:err1}) is 
given by
\begin{equation}
[\bm{\sigma}_{11}^{\ell_{\rm i}}]^2=\sum_{\U_{\rm g_i}} 
w^2_{\rm g_i}\,[\cc_{\ell_{\rm g_i}}(\nu_1,\nu_1) + 
\cc_{\ell_{\rm g_i}}^{\rm N}(\nu_1,\nu_1)]^2\,.
\label{eq:err}
\end{equation}
The two terms in the right-hand side of eq.~\ref{eq:err} are due to the 
cosmic variance and the system noise, respectively. We require the EoR 
21-cm MAPS $\cc_{\ell_{\rm g}}(\nu_1,\nu_2)$, the noise MAPS 
$\cc_{\ell_{\rm g}}^{\rm N}(\nu,\nu)$ and appropriate weights
$w_{\rm g}$ to estimate the errors (eqs.~\ref{eq:err1} and \ref{eq:err}).

We obtain the weights by extremizing the SNR with respect to $w_{\rm g}$ 
with an assumption that the EoR 21-cm MAPS does not vary much within an 
$\ell$-bin and therefore 
$\cc_{\ell_{\rm g_i}}(\nu_1,\nu_2) = \bar{\cl}_{\rm i}(\nu_1,\nu_2)$.  
Note that we consider the variation of the noise 
$\cc_{\ell_{\rm g}}^{\rm N}(\nu,\nu)$ across the grid points 
within a bin. Considering two different frequency channels at $\nu_1$ 
and $\nu_2$, for a particular $\ell$-bin, we can then express the
unnormalized weights in eq.~\ref{eq:err1} as 
\begin{align}
\tilde{w}_{\rm g} = &[\{\cc_{\ell_{\rm g_i}}(\nu_1,\nu_1) + 
\cc_{\ell_{\rm g_i}}^{\rm N}(\nu_1,\nu_1)\}\{\cc_{\ell_{\rm g_i}}(\nu_2,\nu_2) + 
\cc_{\ell_{\rm g_i}}^{\rm N}(\nu_2,\nu_2)\} \nonumber \\
&+\{\bar{\cl}_{\rm g}(\nu_1,\nu_2) + \delta^{\rm K}_{\nu_1\nu_2}\, 
\cc_{\ell_{\rm g}}^{\rm N}(\nu_1,\nu_2)\}^2]^{-1}\,.
\label{eq:w}
\end{align}
This implies that the grid points with higher noise have lower weights 
and contribute less to the estimator. The grid points that are unsampled
during the observation (i.e. $\tau (\U_g) = 0$ and 
$\cc_{\ell_{\rm g}}^{\rm N}(\nu, \nu)=\infty$) have zero weights, hence 
they do not contribute. Using eqs.~\ref{eq:err1} and \ref{eq:w}, we have 
the expression for the error variance 
\begin{equation}
[\bm{\sigma}_{12}^{\ell_{\rm i}}]^2 =\frac{1}{2} \times 
\frac{1}{\sum_{\U_{\rm g_i}} \tilde{w}_{\rm g_i}}\,.
\label{eq:err2}
\end{equation}

We now discuss the behaviour of the error variance
$[\bm{\sigma}_{12}^{\ell_{\rm i}}]^2$ (eq.~\ref{eq:err2}) in two 
different scenarios. The MAPS error variance consists of the cosmic 
variance and the system noise $\cc_{\ell_{\rm g}}^{\rm N}(\nu, \nu)$. 
We see from eq.~\ref{eq:cln} that the noise contribution drops off as 
$\cc_{\ell_{\rm g}}^{\rm N}(\nu, \nu) \propto 1/t_{\rm obs}$ with an 
increase in  observation time. For small observation times, the 
estimated error variance  is thus dominated by the large system 
noise,  and from eq.~\ref{eq:err2}, we have
\begin{equation}
[\bm{\sigma}_{12}^{\ell_{\rm i}}]^2 \simeq \frac{{\mathcal C}^{\rm 0}(\nu_1)
{\mathcal C}^{\rm 0} (\nu_2)+\delta^{\rm K}_{\nu_1\nu_2}[{\mathcal C}^{\rm 0} 
(\nu_1,\nu_2)]^2}{2\times \sum_{\U_{g_{i}}} [\tau(\U_{g_{i}})]^2} \times 
\left(\frac{8\,{\rm hrs}}{t_{\rm obs}}\right)^2\,.
\label{eq:err3}
\end{equation}
In contrast, we have  the other extreme 
$\cc_{\ell_{\rm g}}^{\rm N}(\nu,\nu) \simeq 0$  for very large 
observation times ($t_{\rm obs}\rightarrow \infty$).  In this case, the 
error variance approaches the cosmic variance (CV) limit and we have
\begin{equation}
[\bm{\sigma}_{12}^{\ell_{\rm i}}]^2 \simeq \frac{
\bar{\cl}_{\rm g_i}(\nu_1,\nu_1)\bar{\cl}_{\rm g_i}(\nu_2,\nu_2)
+[\bar{\cl}_{\rm g_i}(\nu_1,\nu_2)]^2}{2 N_{\rm g_i}}\,,
\label{eq:err4}
\end{equation}
where $N_{\rm g_i}$ is the number of sampled grid points in the $i$-th 
bin. 

%%%%%%%%%%%%%%%%%%%%%%%%%%%%%%%%%%%%%%%%%%%%%%%%%%%%%%%%%%%%%%%%%%%%%%%%%

\section{Results}
\label{sec:results}
%%%%%%%%%%%%%%%%%%%%%%%%%%%%%%%%%%%%%%%%%%%%%%%%%%%%%%%%%%%%%%%%%%%%%%%%%
\setlength{\unitlength}{1cm}
\begin{figure*}
\centering
\psfrag{nu2}[c][c][1.2]{$\nu_2 - \nu_{\rm c}$ (MHz)}
\psfrag{nu1}[c][c][1.2]{$\nu_1 - \nu_{\rm c}$ (MHz)}
\psfrag{SNR}[c][c][1.1]{SNR}

\includegraphics[width=0.95\textwidth, angle=0]{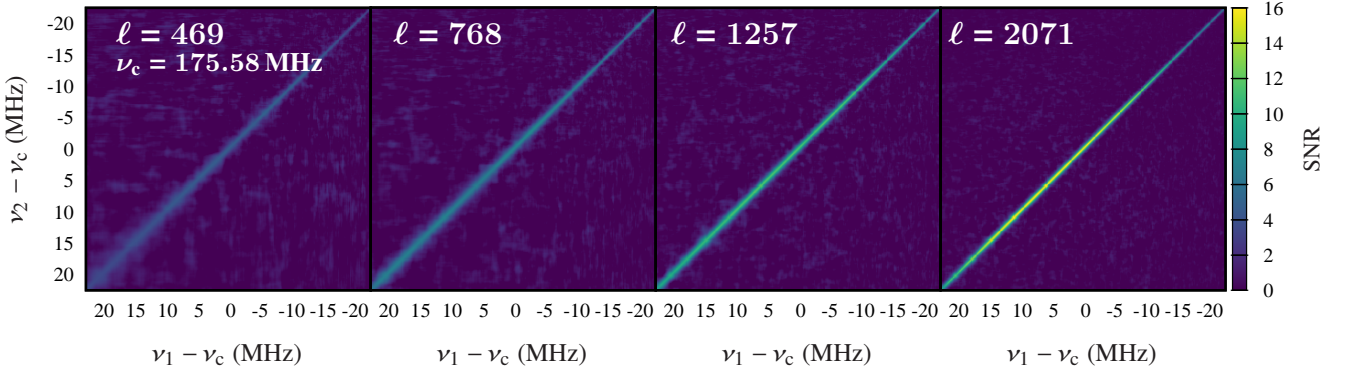}
\put(-15.60,4.2){{\Large \textcolor{white}{\bm{$\ell = 469$}}}}
\put(-11.95,4.2){{\Large \textcolor{white}{\bm{$\ell = 768$}}}}
\put(-8.30,4.2){{\Large \textcolor{white}{\bm{$\ell = 1257$}}}}
\put(-4.65,4.2){{\Large \textcolor{white}{\bm{$\ell = 2071$}}}}
\put(-15.60,3.85){{\large \textcolor{white}{\bm{$\nu_{\rm c}=175.58\,{\rm MHz}$}}}}

\caption{This shows the SNR for MAPS at $\ell = 469$, $768$, $1257$ and 
$2071$ (from left to right, respectively) for $t_{\rm obs}=1024$\,hrs for 
the LC1.}
\label{fig:cl_nu12_0.50}
\end{figure*}
%%%%%%%%%%%%%%%%%%%%%%%%%%%%%%%%%%%%%%%%%%%%%%%%%%%%%%%%%%%%%%%%%%%%%%%%%

%%%%%%%%%%%%%%%%%%%%%%%%%%%%%%%%%%%%%%%%%%%%%%%%%%%%%%%%%%%%%%%%%%%%%%%%%
\setlength{\unitlength}{1cm}
\begin{figure*}
\centering
\psfrag{nu2}[c][c][1.2]{$\nu_2 - \nu_{\rm c}$ (MHz)}
\psfrag{nu1}[c][c][1.2]{$\nu_1 - \nu_{\rm c}$ (MHz)}
\psfrag{SNR}[c][c][1.1]{SNR}

\includegraphics[width=0.95\textwidth, angle=0]{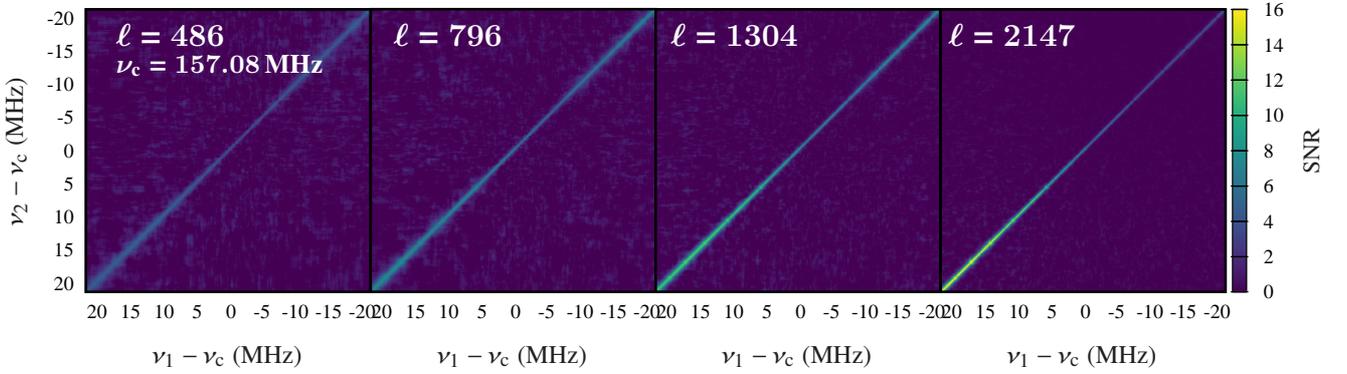}
\put(-15.60,4.2){{\Large \textcolor{white}{\bm{$\ell = 486$}}}}
\put(-11.95,4.2){{\Large \textcolor{white}{\bm{$\ell = 796$}}}}
\put(-8.30,4.2){{\Large \textcolor{white}{\bm{$\ell = 1304$}}}}
\put(-4.65,4.2){{\Large \textcolor{white}{\bm{$\ell = 2147$}}}}
\put(-15.60,3.85){{\large \textcolor{white}{\bm{$\nu_{\rm c}=157.08\,{\rm MHz}$}}}}

\caption{Same as Figure~\ref{fig:cl_nu12_0.50} at $\ell = 486$, $796$, 
$1304$ and $2147$ (from left to right, respectively) for $t_{\rm obs}=1024$ hrs for the LC2.}
\label{fig:cl_nu12_0.75}
\end{figure*}
%%%%%%%%%%%%%%%%%%%%%%%%%%%%%%%%%%%%%%%%%%%%%%%%%%%%%%%%%%%%%%%%%%%%%%%%%

Figures~\ref{fig:cl_nu12_0.50} and~\ref{fig:cl_nu12_0.75} show the SNR 
for measuring the MAPS at the four representative $\ell$ values 
considered here for LC1 and LC2, respectively. For the moderate 
observation time $t_{\rm obs}=1024$\,hrs, we see a correspondence of 
behaviour between the SNR for MAPS and the signal (Figures 
\ref{fig:cl0.50} and \ref{fig:cl0.75}). They both peak along the diagonal 
and fall rapidly away from the diagonal. The previous error estimates 
\citep{morales05,mcquinn06,zaroubi12,datta14,pober14,ewall16,shaw19} 
are restricted to individually analyzing small frequency intervals 
centered at a particular frequency. However, we see that the error 
estimates, as well as the SNR values for MAPS, change with the frequency 
across the bandwidth. We shall discuss this in more detail in the 
following paragraph.

%%%%%%%%%%%%%%%%%%%%%%%%%%%%%%%%%%%%%%%%%%%%%%%%%%%%%%%%%%%%%%%%%%%%%%%%%
\begin{figure*}
\centering
\psfrag{l=469}[c][c][1.]{$\ell=469$}
\psfrag{l=768}[c][c][1.]{$\ell=768$}
\psfrag{l=1257}[c][c][1.]{$\ell=1257$}
\psfrag{l=2071}[c][c][1.]{$\ell=2071$}
\psfrag{signal}[c][c][1.]{\,Signal}
\psfrag{n=16}[c][c][1.]{$128$\,hrs~\,}
\psfrag{n=128}[c][c][1.]{$1024$\,hrs~\,}
\psfrag{n=1250}[c][c][1.]{$10000$\,hrs~~}
\psfrag{n=6250}[c][c][1.]{$50000$\,hrs~~}
\psfrag{cv}[c][c][1.]{~CV}
\psfrag{nu}[c][c][1.2]{$\nu - \nu_{\rm c}$ (MHz)}
\psfrag{cl}[c][c][1.1]{$\Phi^2(\nu, \nu)~({\rm mK})^2$}
\psfrag{diff}[c][c][1.1]{$\Delta \Phi^2/\Phi^2$}

\includegraphics[width=0.99\textwidth, angle=0]{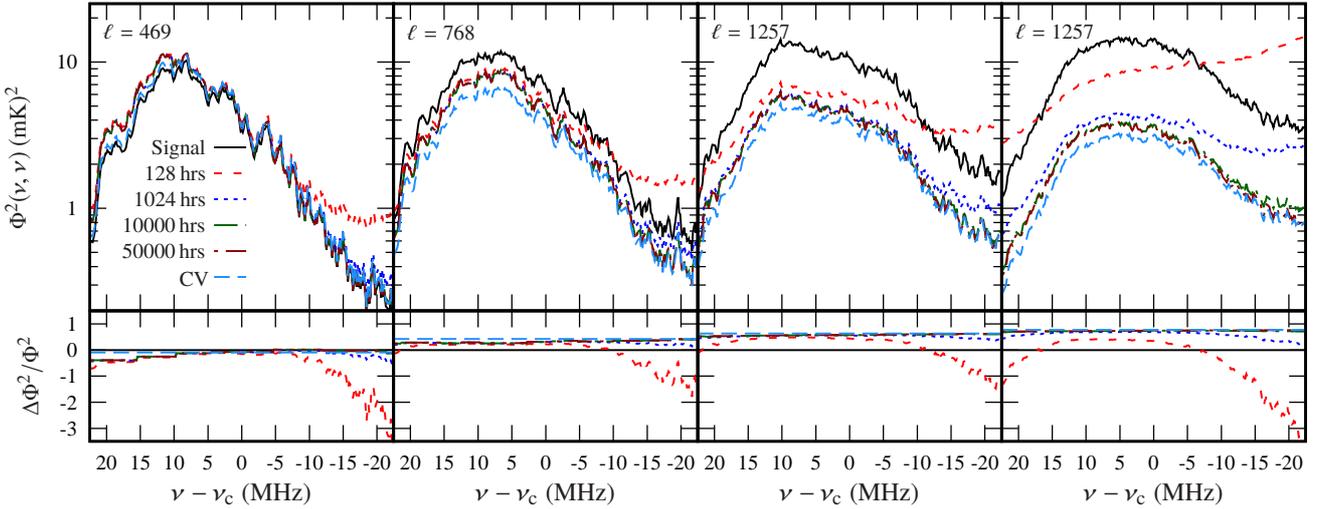}
\caption{This shows the diagonal components of the scale-independent MAPS $\Phi^2(\nu, \nu)$ and the corresponding $5\sigma$ r.m.s. error estimates for the LC1. We consider four different observation times $t_{\rm obs}$. We also show CV which corresponds to $t_{\rm obs} \rightarrow \infty$. The bottom panels show the relative difference $\Delta \Phi^2/\Phi^2 = (\Phi^2(\nu, \nu) - 5\sigma_{11})/\Phi^2(\nu, \nu)$. The positive values in the bottom panels correspond to values that are above the 5$\sigma$ noise level.}
\label{fig:cl_nu11_0.50}
\end{figure*}
%%%%%%%%%%%%%%%%%%%%%%%%%%%%%%%%%%%%%%%%%%%%%%%%%%%%%%%%%%%%%%%%%%%%%%%%%

%%%%%%%%%%%%%%%%%%%%%%%%%%%%%%%%%%%%%%%%%%%%%%%%%%%%%%%%%%%%%%%%%%%%%%%%%
\begin{figure*}
\centering
\psfrag{l=486}[c][c][1.1]{~~$\ell=486$}
\psfrag{l=796}[c][c][1.1]{~~$\ell=796$}
\psfrag{l=1304}[c][c][1.1]{~~$\ell=1304$}
\psfrag{l=2147}[c][c][1.1]{~~$\ell=2147$}
\psfrag{signal}[c][c][1.1]{Signal}
\psfrag{n=16}[c][c][1.1]{$128$\,hrs~\,}
\psfrag{n=128}[c][c][1.1]{$1024$\,hrs~~~}
\psfrag{n=1250}[c][c][1.1]{$10000$\,hrs~~~\,}
\psfrag{n=6250}[c][c][1.1]{$50000$\,hrs~~~\,}
\psfrag{cv}[c][c][1.1]{CV}
\psfrag{nu}[c][c][1.2]{$\nu - \nu_{\rm c}$ (MHz)}
\psfrag{cl}[c][c][1.2]{$\Phi^2(\nu, \nu)~({\rm mK})^2$}
\psfrag{diff}[c][c][1.1]{$\Delta \Phi^2/\Phi^2$}

\includegraphics[width=0.99\textwidth, angle=0]{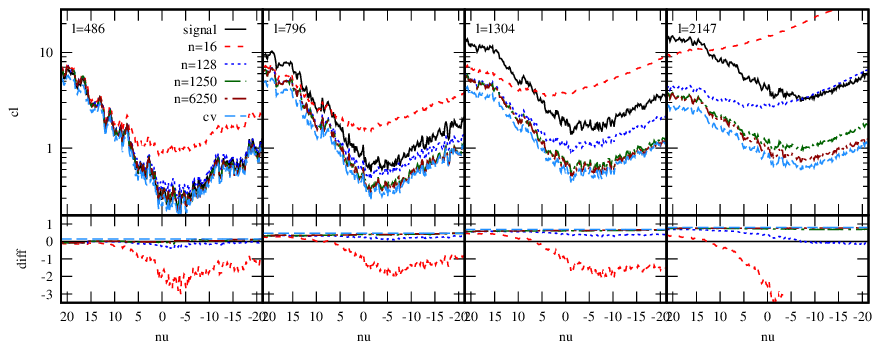}
\caption{Same as Figure~\ref{fig:cl_nu11_0.50} for the LC2.}
\label{fig:cl_nu11_0.75}
\end{figure*}
%%%%%%%%%%%%%%%%%%%%%%%%%%%%%%%%%%%%%%%%%%%%%%%%%%%%%%%%%%%%%%%%%%%%%%%%%

In the subsequent results, we focus on the diagonal elements 
$\nu_1=\nu_2$ of MAPS. Figures~\ref{fig:cl_nu11_0.50} and 
\ref{fig:cl_nu11_0.75} show $\Phi^2(\nu, \nu)$ and the corresponding 
$5\sigma$ r.m.s. error estimates for LC1 and LC2, respectively. 
In these figures, the bottom panels show the relative difference $\Delta \Phi^2/\Phi^2 
= (\Phi^2(\nu, \nu) - 5\sigma_{11})/\Phi^2(\nu, \nu)$. The positive values in the bottom panels correspond to values that are above the 5$\sigma$ noise level. In the 
following analysis, we have considered four different observation times 
$t_{\rm obs}=128$\,hrs, 1024\,hrs, 10000\,hrs, and 50000\,hrs. We also 
show CV, which corresponds to $t_{\rm obs} \rightarrow \infty$ and the 
system noise approaches zero. As discussed above, the cosmic variance 
and the system noise contribute to the total error budget 
(eq.~\ref{eq:err}). Considering the behaviour of r.m.s. error at large 
angular scales, we see that r.m.s. error is not much affected even if 
$t_{\rm obs}$ is increased. Whereas the r.m.s. error decreases as 
$t_{\rm obs}$ is increased at small angular scales. This confirms the 
fact that the cosmic variance dominates the total error at small 
$\ell$ and the system noise contribution dominates at large $\ell$. We also 
see that the r.m.s. error increases with decreasing frequency across 
the bandwidth of our simulations. This is due to the fact that the system 
noise contribution increases (eq.~\ref{eq:cln}) with decreasing frequency. 
Considering Figure~\ref{fig:cl_nu11_0.50}, we see that for any feasible 
$t_{\rm obs}$ a $5\sigma$ detection the MAPS will not be possible 
at $\ell \le 496$. The condition improves at $\ell=796$, where SKA 
will be able to measure the MAPS at $\ge 5\sigma$ confidence over 
$\sim 25$\,MHz frequency band for $t_{\rm obs} \ge 128$\,hrs. $\ell=1257$ 
is a better scenario among the four $\ell$ values, where $5\sigma$ 
detection will be possible roughly across the entire observational 
bandwidth for $t_{\rm obs} \ge 128$\,hrs. Whereas, the frequency 
band allowed for $\ge 5\sigma$ detection reduces at $\ell=2071$ due 
to system noise domination. We find the behaviour in 
Figure~\ref{fig:cl_nu11_0.75} is similar to that in LC1. The optimal 
angular multipole for detection, among the four $\ell$ values, is 
$\ell = 1304$ in Figure~\ref{fig:cl_nu11_0.75}. The difference here 
is that the MAPS signal peaks at one end of the band as compared to 
LC1, where the signal peaks around the centre of the frequency band. 

%%%%%%%%%%%%%%%%%%%%%%%%%%%%%%%%%%%%%%%%%%%%%%%%%%%%%%%%%%%%%%%%%%%%%%%%%
\begin{figure*}
\centering
\psfrag{nu}[c][c][1.2]{$\nu - \nu_{\rm c}$ (MHz)}
\psfrag{l}[c][c][1.2]{$\ell$}
\psfrag{SNR}[c][c][1.1]{SNR}

\includegraphics[width=0.99\textwidth, angle=0]{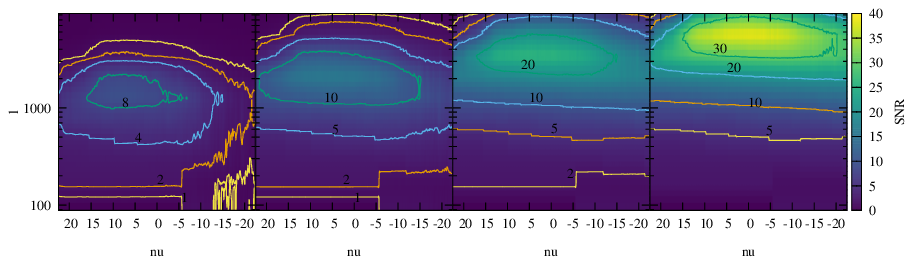}
\caption{This shows the SNR of the diagonal components of MAPS 
$\Phi^2(\nu, \nu)$ as a function $\ell$ at 
$t_{\rm obs} = 128\,{\rm hrs},\, 1024\,{\rm hrs},\, 
10000\,{\rm hrs\,, and\,} 50000\,{\rm hrs}$ 
(from left to right respectively) for the LC1.}
\label{fig:SNR0.50}
\end{figure*}
%%%%%%%%%%%%%%%%%%%%%%%%%%%%%%%%%%%%%%%%%%%%%%%%%%%%%%%%%%%%%%%%%%%%%%%%%

%%%%%%%%%%%%%%%%%%%%%%%%%%%%%%%%%%%%%%%%%%%%%%%%%%%%%%%%%%%%%%%%%%%%%%%%%
\begin{figure*}
\centering
\psfrag{nu}[c][c][1.2]{$\nu - \nu_{\rm c}$ (MHz)}
\psfrag{l}[c][c][1.2]{$\ell$}
\psfrag{SNR}[c][c][1.1]{SNR}

\includegraphics[width=0.99\textwidth, angle=0]{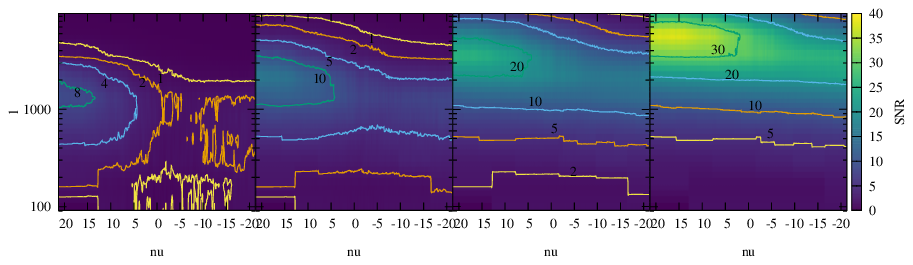}
\caption{Same as Figure~\ref{fig:SNR0.50} for the LC2.}
\label{fig:SNR0.75}
\end{figure*}
%%%%%%%%%%%%%%%%%%%%%%%%%%%%%%%%%%%%%%%%%%%%%%%%%%%%%%%%%%%%%%%%%%%%%%%%%

Figure~\ref{fig:SNR0.50} plots the SNR for the diagonal elements of MAPS
as a function of $\ell$ and frequency for four observation times 
$t_{\rm obs} = 128\,{\rm hrs},\, 1024\,{\rm hrs},\,10000\,{\rm hrs}$, and 
$50000\,{\rm hrs}$ (from left to right, respectively) for LC1 simulation. 
It also shows various contours corresponding to different SNR values. We 
see that the SNR peaks at intermediate scales corresponding to $\ell$ of 
a few thousand. This is because the cosmic variance dominates the total 
error at large scales whereas the system noise dominates at small scales. 
Since the cosmic variance part is independent of the total observing 
time, the SNR at large scales does not improve by increasing the total 
observing time. However, the SNR at small scales (large $\ell$) increases 
with the total observing time. Consequently, the scale at which the SNR 
peaks moves towards the higher $\ell$ values. The SNR drops at lower 
frequencies (higher redshifts) because of the rise in the system 
temperature. We find similar behaviour in Figure~\ref{fig:SNR0.75} which 
shows results for LC2 simulation. The only difference is that the SNR 
is maximum at the highest frequency (lowest redshift) explored here. 
This is because the power spectrum in this simulation is maximum at the 
highest frequency unlike the LC1 simulation where the power spectrum 
peaks at some intermediate frequency.  

%%%%%%%%%%%%%%%%%%%%%%%%%%%%%%%%%%%%%%%%%%%%%%%%%%%%%%%%%%%%%%%%%%%%%%%%%

\section{Effects of foregrounds}
\label{sec:foreground}
Foregrounds pose a major challenge even for the statistical detection of the EoR 21-cm signal. Two main approaches have been proposed in the literature to tackle the foreground problem. One of them is the {\it foreground removal} \citep{Morales2006,ali08,Harker2009,Bonaldi2015,Pober2016,Mertens2018,Mertens2020}, in which the foreground is modelled and removed from the observed 21-cm signal. The forecast of this analysis till now has assumed that the foregrounds have been completely removed, and we refer this as the `Optimistic' scenario \citep{SumanChatterjee_2019,shaw19}.

The other technique is termed as {\it foreground avoidance}. The foreground contamination is found to be restricted within a wedge shaped region in the ($\kk_{\perp}, k_{\parallel}$) plane \citep{Adatta2010} with the wedge boundary defined by
\begin{equation}
k_{\parallel}=\left[ \frac{r_{\rm c} \sin{(\theta_{\rm L}})}{r_{\rm c}^{\prime}\nu_{\rm c}} \right]k_{\perp} \,,
\end{equation}
where $r_{\rm c}$ is the comoving distance corresponding to the central frequency $\nu_{\rm c}$, $r_{\rm c}^{\prime}=\frac{\partial r_{\rm c}}{\partial \nu}\mid_{\nu_{\rm c}}$ and $\theta_{\rm L}$ is the angle on the sky with respect to the zenith from which the foregrounds contaminate
the EoR 21-cm signal. The value of $\theta_{\rm L}$ and hence the slope of the wedge is
determined by the level of foreground contamination \citep{Morales2012,hassan2019}. The region
outside the foreground wedge, the `EoR Window', is utilised for estimating the EoR 21-cm 3D
power spectrum $P(\kk)$ \citep{pober13,Kerrigan2018}. The upper limit of the wedge boundary is
set by the horizon for which $\theta_{\rm L}=90^{\circ}$. We refer to this case as the
`Pessimistic' scenario \citep{SumanChatterjee_2019,shaw19}.

\citet{ghosh11} and \citet{choudhuri16} have shown that tapering the sky response in
telescope's field of view restricts the $\theta_{\rm L}$ to an angle smaller than the horizon
limit. In this analysis, we consider two different tapering situations which assume
$\theta_{\rm L}=3 \times ({\rm FWHM}/2)$ and $\theta_{\rm L}=9 \times ({\rm FWHM}/2)$. Here,
FWHM is the Full Width Half Maximum of the SKA-Low primary beam. We refer these two tapered
cases as the `Mild' and `Moderate' scenarios respectively \citep{SumanChatterjee_2019,shaw19}.

As mentioned earlier, our estimator $\cl(\nu_1,\nu_2)$ does not assume that the signal is ergodic and periodic along the LoS direction. Therefore, it is not straightforward to consider the foreground wedge in our analysis as this explicitly assumes the signal to be ergodic and periodic along LoS direction. Following \citet{mondal18}, we define $\cl^{{\rm EP}}(\nu_1,\nu_2)$ which is the ergodic\,(E) and periodic\,(P) component of $\cl(\nu_1,\nu_2)$. The $\cl^{ {\rm EP}}(\nu_1,\nu_2)$ is estimated from the measured $\cl(\nu_1,\nu_2)$ by imposing the conditions $\cl^{{\rm EP}}(\nu_1,\nu_2)=\cl^{{\rm EP}}(\mid \nu_1-\nu_2 \mid)=\cl^{{\rm EP}}(\Delta \nu)$ (ergodicity) and $\cl^{{\rm EP}}(\Delta \nu) =\cl^{{\rm EP}}(B-\Delta \nu)$ (periodicity). Under these assumptions, we can estimate $\cl^{{\rm EP}}(\Delta \nu)$ using eq.~21 of \citet{datta07a} as:
\begin{equation}
\cl^{\rm EP}(\Delta\nu)=\frac{1}{\pi r_{\rm c}^2} \int^{\infty}_0 dk_{\parallel} \cos{(k_{\parallel} r^{\prime}_{\rm c} \Delta\nu)}~
P(k_{\perp},k_{\parallel})\,.
\label{eq:pkcl}
\end{equation}
%\begin{equation}
%\cl^{\rm EP}(\Delta\nu)=(r_{\rm c}^2 r^{\prime}_{\rm c} B)^{-1}
%\sum_{k_{\parallel}} {\rm e}^{i k_{\parallel}  r^{\prime}_{\rm c} \Delta\nu}~
%P(k_{\perp},k_{\parallel})\,,
%\label{eq:pkcl}
%\end{equation}
%we can estimate $P(\kk_{\perp}, k_{\parallel})$ using  
%\begin{equation}
%P(k_{\perp},\,k_{\parallel})= r_{\rm c}^2\,r^{\prime}_{\rm c} \int d (\Delta \nu) \, e^{-i  k_{\parallel} r^{\prime}_{\rm c} \Delta  \nu}\, \cl^{\rm EP}(\Delta \nu)\,.
%\label{eq:cl_Pk}
%\end{equation}

We compute the foreground `wedge' boundary in ($\kk_\perp,k_\parallel$) plane and discard the signal contained within it. A sharp cut-off at the wedge boundary introduces ripples in the estimated $\cl^{{\rm EP}}(\Delta \nu)$ while performing the Fourier transform (eq.~\ref{eq:pkcl}). In order to avoid this issue, we multiply the $P(k_\perp,k_\parallel)$ with a Butterworth filter of order 16 
\begin{equation}
\mathcal{B}(k_{\perp}, k_{\parallel}) = \frac{1}{\sqrt{1+\left(\frac{r_{\rm c}^{\prime} \, \nu_{\rm c} \, k_{\parallel}}{r_{\rm c} \sin{(\theta_{\rm L})}\, k_{\perp}}\right)^{32}}} \,.
\end{equation}
The value of the Butterworth function is $1$ within wedge and it decays sharply but continuously to zero outside the wedge. The value is $1/\sqrt{2}$ at the wedge boundary. By changing the order of the Butterworth function one can control its steepness and also the amount of foreground spill into the EoR window \citep{SumanChatterjee_2020}. We then take Fourier transform of $P(\kk_{\perp}, k_{\parallel})\,\mathcal{B}(k_{\perp}, k_{\parallel})$ to get the foreground contaminated $\cl^{{\rm EP}}(\Delta \nu)$ (eq.~\ref{eq:pkcl}). Finally, we use the `Foreground Avoidance' MAPS $\cl^{\rm FA}(\nu_1,\nu_2)= \cl(\nu_1,\nu_2)-\cl^{\rm EP}(\Delta \nu)$ for the MAPS SNR predictions. This leads to subtraction of a part from MAPS at each frequency at any particular $\ell$ value.

%%%%%%%%%%%%%%%%%%%%%%%%%%%%%%%%%%%%%%%%%%%%%%%%%%%%%%%%%%%%%%%%%%%%%%%%%
\begin{figure*}
\centering
\psfrag{nu}[c][c][1.2]{$\nu - \nu_{\rm c}$ (MHz)}
\psfrag{l}[c][c][1.2]{$\ell$}
\psfrag{SNR}[c][c][1.1]{SNR}

\includegraphics[width=0.99\textwidth, angle=0]{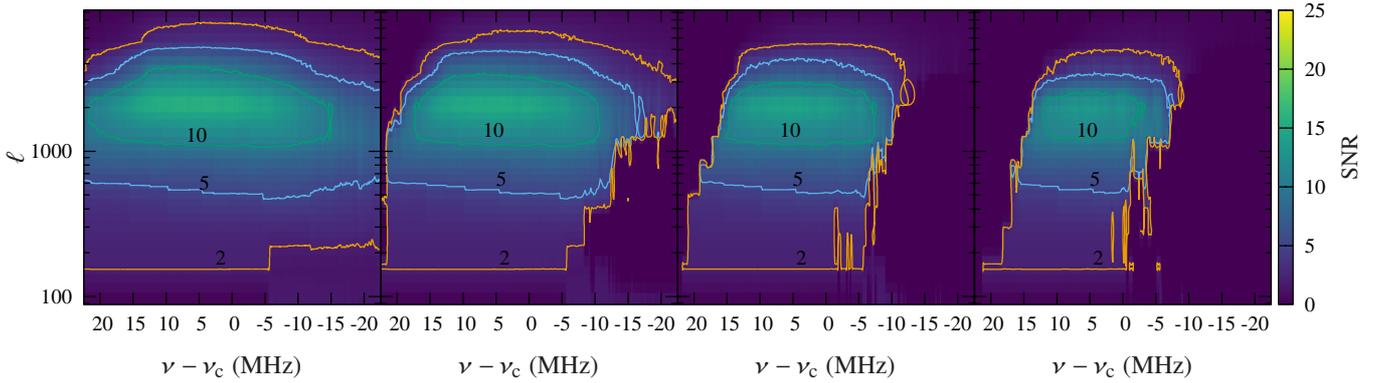}
\caption{This shows the SNR of the diagonal components of MAPS $\Phi^2(\nu, \nu)$ as a function $\ell$ at $t_{\rm obs} = 1024\,{\rm hrs}$ for `Optimistic', `Mild', `Moderate' and `Pessimistic' (from left to right respectively) for the LC1.}
\label{fig:Fore_SNR0.50}
\end{figure*}
%%%%%%%%%%%%%%%%%%%%%%%%%%%%%%%%%%%%%%%%%%%%%%%%%%%%%%%%%%%%%%%%%%%%%%%%%

%%%%%%%%%%%%%%%%%%%%%%%%%%%%%%%%%%%%%%%%%%%%%%%%%%%%%%%%%%%%%%%%%%%%%%%%%
\begin{figure*}
\centering
\psfrag{nu}[c][c][1.2]{$\nu - \nu_{\rm c}$ (MHz)}
\psfrag{l}[c][c][1.2]{$\ell$}
\psfrag{SNR}[c][c][1.1]{SNR}

\includegraphics[width=0.99\textwidth, angle=0]{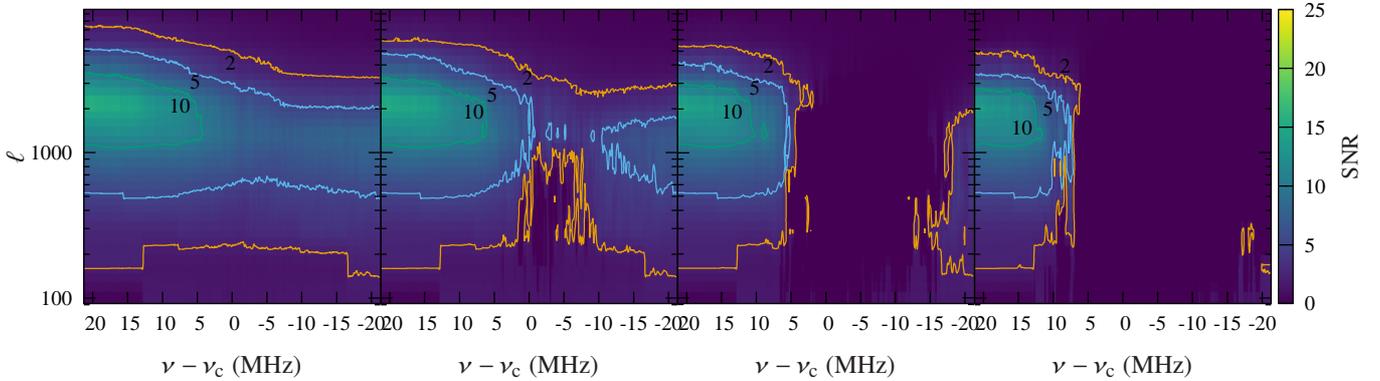}
\caption{Same as Figure~\ref{fig:Fore_SNR0.50} for the LC2.}
\label{fig:Fore_SNR0.75}
\end{figure*}
%%%%%%%%%%%%%%%%%%%%%%%%%%%%%%%%%%%%%%%%%%%%%%%%%%%%%%%%%%%%%%%%%%%%%%%%%

Figures~\ref{fig:Fore_SNR0.50} and \ref{fig:Fore_SNR0.75} show the SNR predictions for the diagonal components of MAPS $\Phi^2 (\nu,\nu)$ as a function $\ell$ at $t_{\rm obs}=1024$\,hrs for LC1 and LC2, respectively. In these figures, the four panels show the predictions for the `Optimistic', `Mild', `Moderate' and `Pessimistic' scenarios starting from the left to the right respectively. The first obvious point is that the SNR gradually decreases from the Optimistic to the Pessimistic scenario. This happens due to the fact that more number of $k_{\parallel}$ modes are being discarded from the Optimistic to the Pessimistic scenario due to increase in foreground wedge. Note that the extent of the foreground wedge increases at lower $\nu_{\rm c}$ due to the factor $r_{c}/r^{\prime}_{c}\nu_{c} \sim \sqrt{1420/\nu_{\rm c}}$ and $\theta_{\rm L}$ as well \citep{shaw19}. As a consequence, we see the effects of the foreground wedge is more for LC2 (Figure~\ref{fig:Fore_SNR0.75}) as compared to LC1 (Figure~\ref{fig:Fore_SNR0.50}). Considering the region where the SNR exceeds the value $5$ for LC1, we see the detection of MAPS is possible over the full available band-width at $\ell \sim 2000$ for Optimistic scenario. However, the band-width for a $5\sigma$ detection reduces to $\sim 40$\,MHz, $\sim 25$\,MHz and  $\sim 20$\,MHz for Mild, Moderate and Pessimistic scenarios, respectively. We find qualitatively similar behaviour in Figure~\ref{fig:Fore_SNR0.75} which shows the results for LC2 simulation. As mentioned above, the impacts of foreground avoidance is more for LC2 box for being centered at a smaller frequency (higher redshift). Note that this analysis is particularly valid where the signal is ergodic and periodic along the LoS. However, at the moment we do not have a clear picture of how to tackle the foreground problem in the context of MAPS. This is a problem worthy of detailed investigations in future.

%%%%%%%%%%%%%%%%%%%%%%%%%%%%%%%%%%%%%%%%%%%%%%%%%%%%%%%%%%%%%%%%%%%%%%%%%

\section{Discussion and Conclusions}
\label{sec:discuss}
Several observational efforts are underway to detect the EoR 21-cm power 
spectrum $P(\kk)$ using the presently operating radio-interferometers 
across the globe. One of the key science goals of the future telescope 
SKA-Low is to measure the spherically averaged 3D EoR power spectrum 
$P(k)$. The definition of the spherically averaged 3D power spectrum 
makes use of the assumption that the signal is ergodic and periodic 
in all three spatial directions. However, the LC effect breaks both
ergodicity and periodicity along the LoS of the observer. The problem is 
particularly severe during EoR where the mean \HI fraction $\xb$ changes 
rapidly with redshift and this affects large bandwidth observations with
different telescopes \citep{mondal19}. The spherically averaged 3D power 
spectrum $P(k)$ can no longer, therefore, be regarded as the correct 
estimator to quantify the second order statistics of the EoR 21-cm signal 
\citep{mondal18}, and any estimation using this may lead to a biased 
estimate for the statistics of the signal \citep{trott16}. As an 
alternative to $P(k)$, we have used the MAPS $\cl(\nu_1, \nu_2)$ to 
quantify the two-point statistics of the EoR 21-cm signal. This does 
not assume the signal to be ergodic and periodic along the LoS. The 
only assumption here is that the signal is statistically homogeneous 
and isotropic in different directions on the observing plane of the sky. 
In this work, we make predictions on the SNR for measuring the EoR 
21-cm MAPS using the future radio interferometer SKA-Low.

The sensitivity of any instrument to the measurement of the EoR 21-cm 
MAPS is limited by the errors, a part of which is inherent to the signal 
itself (cosmic variance), and the other part arises due to the system 
noise (external contamination). The EoR 21-cm signal is expected to be a 
highly non-Gaussian field \citep{bharadwaj05a,mondal15,majumdar18}. 
The effects of this inherent non-Gaussianity play a significant role in 
the error estimates of the two-point correlation functions of the signal 
\citep{mondal16, mondal17, shaw19}. The non-Gaussianity of the signal introduces 
non-zero trispectrum contribution in the error covariance of the 21-cm MAPS. However in 
this work, we assume that the observed 21-cm MAPS error covariance is well
approximated by that of a Gaussian field predictions and ignored the trispectrum
contribution for the simplicity of computations. We have used a 3D 
radiative transfer code \textsc{\small C$^2$-Ray} to generate the EoR
21-cm LC signal and incorporated observational effects like the system noise and the
array baseline distribution to predict the prospects of observing the 
bin-averaged MAPS using SKA-Low. We have considered two observations 
{\bf LC1} centered at $\nu_{\rm c}=175.58$\,MHz, which corresponds to 
$\xb \approx 0.50$, and {\bf LC2} centered at $\nu_{\rm c}=157.08$\,MHz, 
corresponding to $\xb \approx 0.75$. We also present a detailed 
theoretical framework to quantify and interpret the error estimates for 
the MAPS incorporating the system noise.

For moderate observation times, we have seen that the r.m.s. error $\sigma$
scales as $1/t_{\rm obs}$ and consequently we have ${\rm SNR} \propto 
t_{\rm obs}$. In this case, we have found similar behaviour between the 
signal and the SNR for MAPS (Figures~\ref{fig:cl0.50}, \ref{fig:cl0.75}, 
\ref{fig:cl_nu12_0.50} and \ref{fig:cl_nu12_0.75}). They both peak along 
the diagonal $\nu_1=\nu_2$ and fall rapidly away from the diagonal. For 
further analysis, we have focused on the diagonal elements of the MAPS. 
We have found that the error predictions for MAPS are not much affected by 
the choice of $t_{\rm obs}$ at large angular scales. This is due to the 
fact that cosmic variance dominates the total error budget at small $\ell$. 
We have also found that the r.m.s. error decreases as $t_{\rm obs}$ is 
increased at small angular scales. This is because the system noise 
dominates the total error at large $\ell$. We have found that a $5\sigma$ 
detection of MAPS will not be possible with SKA-Low at $\ell \le 496$ for 
LC1, and at $\ell \le 486$ for LC2. Although, we have found that the SKA 
will be able to measure the MAPS at $\ge 5\sigma$ confidence roughly 
across the $\sim 44$\,MHz observational bandwidth at $\ell \sim 1300$ with 
$t_{\rm obs} \ge 128$~hrs. Whereas, the frequency band allowed for
$\ge 5\sigma$ detection reduces at higher values of $\ell$ due to system 
noise domination. We have noted that the r.m.s. error increases with 
decreasing frequency across the bandwidth of our simulations. This is due 
to the fact that the system noise contribution increases (eq.~\ref{eq:cln}) 
with decreasing frequency. We have extended our analysis to study how the 
SNR for the diagonal elements of 21-cm MAPS changes with $\ell$ 
(Figures~\ref{fig:SNR0.50} and \ref{fig:SNR0.75}). Note that the system 
noise contribution, at a fixed $\ell$, decreases with as $t_{\rm obs}$ 
increases but the cosmic variance remains unchanged. However, the cosmic 
variance dominates the total error at larger values of $\ell$ for a fixed 
$t_{\rm obs}$. This interplay between the system noise and cosmic variance 
(as a function of $t_{\rm obs}$) causes the peak of the SNR for MAPS to 
shifts toward larger values of $\ell$ as $t_{\rm obs}$ is increased 
(Figures~\ref{fig:SNR0.50} and \ref{fig:SNR0.75}).

In Section~\ref{sec:results}, we have assumed that the foregrounds are completely removed from the signal. We refer this as the `Optimistic' scenario. In Section~\ref{sec:foreground}, we have then attempted to quantitatively address the effects of foregrounds on MAPS detectability forecast. The foreground contamination is found to be restricted within a wedge shaped region in the ($\kk_{\perp}, k_{\parallel}$) plane \citep{Adatta2010} and the region outside the foreground wedge is utilised for estimating the EoR 21-cm 3D power spectrum $P(\kk)$ in Foreground avoidance technique \citep{Morales2012,pober13,Kerrigan2018,hassan2019}. We have consider three foreground scenarios `Mild', `Moderate' and `Pessimistic' respectively. We have found that the SNR gradually decreases from the Optimistic to the Pessimistic scenario. The band-width for a 5$\sigma$ detection at $\ell \sim 2000$ reduces from full available band-width to $\sim 40$\,MHz, $\sim 25$\,MHz and $\sim 20$\,MHz for Mild, Moderate and Pessimistic scenarios, respectively for LC1. We have found qualitatively similar behaviour for LC2. However, the impacts of foreground avoidance is more for LC2 for being centered at a smaller frequency (higher redshift).

In conclusion, our study indicates that the EoR 21-cm MAPS, which is 
directly related to the correlations between the visibilities measured 
in radio-interferometric observation, can be measured at a confidence 
level of $5\sigma$ or more at angular multipole $\ell \sim 1300$ for 
$t_{\rm obs} \ge 128$\,hrs across $\sim 44$~MHz observational bandwidth 
using SKA-Low. The framework presented in this paper is general and can 
be applied to any radio-interferometer given the array baseline distribution.

%--------------------------------------------------------------------

\section*{Acknowledgements}
This work was supported by the Science and Technology Facilities Council
[grant numbers ST/F002858/1 and ST/I000976/1] and the Southeast Physics
Network~(SEPNet). We acknowledge that the results in this paper have been 
achieved using the PRACE Research Infrastructure resources Curie based at 
the Tr\`es Grand Centre de Calcul (TGCC) operated by CEA near Paris, France 
and Marenostrum based in the Barcelona Supercomputing Center, Spain. Time 
on these resources was awarded by PRACE under PRACE4LOFAR grants 2012061089 
and 2014102339 as well as under the Multi-scale Reionization grants 
2014102281 and 2015122822. The authors gratefully acknowledge the Gauss 
Centre for Supercomputing e.V. (www.gauss-centre.eu) for funding this project 
by providing computing time through the John von Neumann Institute for 
Computing (NIC) on the GCS Supercomputer JUWELS at J\"ulich Supercomputing 
Centre (JSC). Some of the numerical computations were done on the Apollo 
cluster at The University of Sussex.

%--------------------------------------------------------------------

\bibliographystyle{mnras} 
\bibliography{refs}

%--------------------------------------------------------------------
\appendix
\section{MAPS error covariance}
\label{a1}
The error covariance of the MAPS measured at the $i$-th and the $j$-th bins can be written as
\begin{equation}
\begin{aligned}
\bm{X}^{\ell_{\rm i}\ell^{\prime}_{\rm j}}_{12,34}& =\langle [\Hat{\cc}_{\ell_{\rm i}}^t(\nu_1,\nu_2) - \Bar{\cc}_{\ell_{\rm i}}^t(\nu_1,\nu_2)] [\Hat{\cc}_{\ell^{\prime}_{\rm j}}^t(\nu_3,\nu_4) - \Bar{\cc}_{\ell^{\prime}_{\rm j}}^t(\nu_3,\nu_4)]\rangle \\
&=\langle \Hat{\cc}_{\ell_{\rm i}}^t(\nu_1,\nu_2) \Hat{\cc}_{\ell^{\prime}_{\rm j}}^t(\nu_3,\nu_4)\rangle - \Bar{\cc}_{\ell_{\rm i}}^t(\nu_1,\nu_2) \Bar{\cc}_{\ell^{\prime}_{\rm j}}^t(\nu_3,\nu_4).
\end{aligned}
\label{eq:a1_5}
\end{equation}
Using eq.~\ref{eq:est2}, the first term on the right side of the 
eq.~(\ref{eq:a1_5}) can be expressed as
\begin{equation}
\begin{aligned}
&\langle \Hat{\cc}_{\ell_{\rm i}}^t(\nu_1,\nu_2)\Hat{\cc}_{\ell^{\prime}_{\rm j}}^t(\nu_3,\nu_4)\rangle = \frac{1}{4\Omega^2} \sum_{{\bm \U}} \sum_{{\bm \U^\prime}} w({\U}) w({\U^\prime}) \times \\ &[\langle \Tilde{T}^t_{\rm b2}({\U},\nu_1) \Tilde{T}^t_{\rm b2}(-{\U},\nu_2) \Tilde{T}^t_{\rm b2}({\U^\prime},\nu_3) \Tilde{T}^t_{\rm b2}(-{\U^\prime},\nu_4) \rangle \\ &+ \langle \Tilde{T}^t_{\rm b2}({\U},\nu_1) \Tilde{T}^t_{\rm b2}(-{\U},\nu_2) \Tilde{T}^t_{\rm b2}(-{\U^\prime},\nu_3) \Tilde{T}^t_{\rm b2}({\U^\prime},\nu_4) \rangle \\ &+ \langle \Tilde{T}^t_{\rm b2}(-{\U},\nu_1) \Tilde{T}^t_{\rm b2}({\U},\nu_2) \Tilde{T}^t_{\rm b2}({\U^\prime},\nu_3) \Tilde{T}^t_{\rm b2}(-{\U^\prime},\nu_4) \rangle \\ &+ \langle \Tilde{T}^t_{\rm b2}(-{\U},\nu_1) \Tilde{T}^t_{\rm b2}({\U},\nu_2) \Tilde{T}^t_{\rm b2}(-{\U^\prime},\nu_3) \Tilde{T}^t_{\rm b2}({\U^\prime},\nu_4) \rangle]~.
\end{aligned}
\label{eq:a1_6}
\end{equation}
Using $\TTb^{\rm t}(\U,\nu) = \TTb(\U,\nu) + \TTb^{\rm N}(\U,\nu)$ and 
eq.~\ref{eq:clsum}, the first ensemble average in eq.~(\ref{eq:a1_6}) 
can be arranged as
\begin{equation}
\begin{aligned}
&\langle \Tilde{T}^t_{\rm b2}({\U},\nu_1) \Tilde{T}^t_{\rm b2}(-{\U},\nu_2) \Tilde{T}^t_{\rm b2}({\U^\prime},\nu_3) \Tilde{T}^t_{\rm b2}(-{\U^\prime},\nu_4) \rangle = \\
&\Omega^2 [\{\cc_{\ell}(\nu_1,\nu_2)\cc_{\ell^\prime}(\nu_3,\nu_4) + \delta^{\rm K}_{\nu_3\nu_4} \cc_{\ell}(\nu_1,\nu_2)\cc^{\rm N}_{\ell^\prime}(\nu_3,\nu_4) \\ 
&+ \delta^{\rm K}_{\nu_1\nu_2} \cc^{\rm N}_{\ell}(\nu_1,\nu_2) \cc_{\ell^\prime}(\nu_3,\nu_4) + \delta^{\rm K}_{\nu_1\nu_2} \delta^{\rm K}_{\nu_3\nu_4} \cc^{\rm N}_{\ell}(\nu_1,\nu_2) \cc^{\rm N}_{\ell^\prime}(\nu_3,\nu_4)\} \\
&+ \delta^{\rm K}_{\U - \U^\prime,0} \{\cc_{\ell}(\nu_1,\nu_4) \cc_{\ell^\prime}(\nu_2,\nu_3)+ \delta^{\rm K}_{\nu_2\nu_3} \cc_{\ell}(\nu_1,\nu_4) \cc^{\rm N}_{\ell^\prime}(\nu_2,\nu_3) \\ 
&+ \delta^{\rm K}_{\nu_1\nu_4} \cc^{\rm N}_{\ell}(\nu_1,\nu_4) \cc_{\ell^\prime}(\nu_2,\nu_3) + \delta^{\rm K}_{\nu_1\nu_4} \delta^{\rm K}_{\nu_2\nu_3} \cc^{\rm N}_{\ell}(\nu_1,\nu_4) \cc^{\rm N}_{\ell^\prime}(\nu_2,\nu_3)\} \\
&+ \delta^{\rm K}_{\U+\U^\prime,0} \{\cc_{\ell}(\nu_1,\nu_3) \cc_{\ell^\prime}(\nu_2,\nu_4) + \delta^{\rm K}_{\nu_2\nu_4} \cc_{\ell}(\nu_1,\nu_3) \cc^{\rm N}_{\ell^\prime}(\nu_2,\nu_4) \\ 
&+ \delta^{\rm K}_{\nu_1\nu_3} \cc^{\rm N}_{\ell}(\nu_1,\nu_3) \cc_{\ell^\prime}(\nu_2,\nu_4) + \delta^{\rm K}_{\nu_1\nu_3} \delta^{\rm K}_{\nu_2\nu_4} \cc^{\rm N}_{\ell}(\nu_1,\nu_3) \cc^{\rm N}_{\ell^\prime}(\nu_2,\nu_4)\}]~,
\end{aligned}
\label{eq:a1_7}
\end{equation}
where~$\delta^{\rm K}_{\U - \U^\prime,0}$ and 
$\delta^{\rm K}_{\U + \U^\prime,0}$ are Kronecker's delta. Here, 
we have considered that $\TTb(\U,\nu)$ correlates at same baselines 
$\U$ and $\TTb^{\rm N}(\U,\nu)$ correlates at same $\U$ and same 
frequency $\nu$. We obtain this expression after ignoring the non-Gaussianity of the EoR 21-cm signal. However the inherent non-Gaussianity gives rise to a non-zero four point connected term (trispectrum; see eq. 2 of \citealt{adhikari2019}) in the cosmic variance of MAPS. The non-zero trispectra will introduce additional term in eq. (\ref{eq:a1_7}) which will eventually increase the variance and also introduce correlations between errors in the estimated MAPS. We finally ignore trispectrum contribution in our error analysis following the discussion in Section~\ref{sec:error}.

The MAPS estimations are restricted to the upper half of the baseline distribution. Hence, $\delta^{\rm K}_{\U + \U^\prime,0}=0$ and eq.~\ref{eq:a1_7} reduces to
\begin{equation}
\begin{aligned}
\langle \Tilde{T}^t_{\rm b2}({\U},\nu_1) \Tilde{T}^t_{\rm b2}(-{\U},\nu_2) & \Tilde{T}^t_{\rm b2}({\U^\prime},\nu_3) \Tilde{T}^t_{\rm b2}(-{\U^\prime},\nu_4) \rangle = \\
&\Omega^2 [\cc^t_{\ell}(\nu_1,\nu_2) \cc^t_{\ell^\prime}(\nu_3,\nu_4) \\
&+ \delta^{\rm K}_{\U - \U^\prime,0} ~ \cc^t_{\ell}(\nu_1,\nu_4) \cc^t_{\ell^\prime}(\nu_2,\nu_3)]~.
 \end{aligned}
\label{eq:a1_8}
\end{equation}
Similarly, one can write down the other three ensemble averages of 
eq.~(\ref{eq:a1_6}) by permuting the frequency indices in 
eq.~(\ref{eq:a1_8}). 
%The second term on the right hand side of eq. (\ref{eq:a1_5}) when expanded gives
%\begin{equation}
%\begin{aligned
%&\Bar{\cc}_{\ell_{i}}^t(\nu_1,\nu_2)\Bar{\cc}_{\ell_{j}}^t(\nu_3,\nu_4)\\= &\frac{1}{4} \sum_{{\U}} \sum_{{\U^\prime}} w({\U}) w({\U^\prime}) \cc_{{\bm \ell}}^t(\nu_1,\nu_2)\cc_{{\bm \ell^\prime}}^t(\nu_3,\nu_4)\\
%= &\frac{1}{4} \sum_{{\U}} \sum_{{\U^\prime}} w({\U}) w({\U^\prime}) \times \\ &[\cc_{{\bm \ell}}(\nu_1,\nu_2)\cc_{{\bm \ell^\prime}}(\nu_3,\nu_4)+ \cc_{{\bm \ell}}(\nu_1,\nu_2)\cc_{{\bm \ell^\prime}}^{\rm N}(\nu_3,\nu_4)\delta^{\rm K}_{\nu_3\nu_4} \\
%&+ \cc_{{\bm \ell}}^{\rm N}(\nu_1,\nu_2)\cc_{{\bm \ell^\prime}}(\nu_3,\nu_4)\delta^{\rm K}_{\nu_1\nu_2}+\cc_{{\bm \ell}}^{\rm N}(\nu_1,\nu_2)\cc_{{\bm \ell^\prime}}^{\rm N}(\nu_3,\nu_4)\delta^{\rm K}_{\nu_1\nu_2}\delta^{\rm K}_{\nu_3\nu_4}]~.
%\end{aligned}
%\label{eq:a1_9}
%\end{equation}
Combining eq.~(\ref{eq:a1_5}) and (\ref{eq:a1_8}) with the other ensemble averages, we write the error covariance in compact form
\begin{equation}
\begin{aligned}
\bm{X}^{\ell_{\rm i}\ell^{\prime}_{\rm j}}_{12,34}=\frac{1}{2} \delta^{\rm K}_{\U-\U^\prime,0} \sum_{{\U}} \sum_{{\U^\prime}} w({\U}) w({\U^\prime})
&\, [\cc_{\ell_{\rm i}}^t(\nu_1,\nu_3)\cc_{\ell^{\prime}_{\rm j}}^t(\nu_2,\nu_4) \\
&+\cc_{\ell_{\rm i}}^t(\nu_1,\nu_4)\cc_{\ell^{\prime}_{\rm j}}^t(\nu_2,\nu_3)]~.
\end{aligned}
\label{eq:a1_10}
\end{equation}
Finally, exploiting the Kronecker's delta, we obtain
\begin{equation}
\cov_{12,34}=\frac{1}{2} \sum_{\U} w^2 [\cc_{\bm \ell}^t(\nu_1,\nu_3)\cc_{\bm \ell}^t(\nu_2,\nu_4) +\cc_{\bm \ell}^t(\nu_1,\nu_4)\cc_{\bm \ell}^t(\nu_2,\nu_3)]~.
\end{equation}
%--------------------------------------------------------------------

\vfill
\bsp

\label{lastpage}

\end{document}